\documentclass[aps,prl,twocolumn,superscriptaddress,longbibliography]{revtex4-1}
\usepackage{amsthm}
\usepackage[utf8]{inputenc}
\usepackage{amsmath,amssymb,amsmath,amsthm,bm,graphicx,xcolor}
\usepackage{natbib}
\usepackage[breaklinks]{hyperref}
\usepackage{color}
\usepackage{tabularx}
\usepackage{epsfig}
\usepackage{amssymb}
\usepackage{comment}
\usepackage{graphicx}
\usepackage{xcolor}
\usepackage{ytableau}
\usepackage{wasysym}
\usepackage{nicefrac}
\usepackage{dcolumn}
\usepackage{epstopdf}
\usepackage{subfigure}
\usepackage{psfrag}
\usepackage{bbold}
\hypersetup{colorlinks=true}
\usepackage{lineno}
\usepackage{mathtools}
\usepackage{xfrac}

\usepackage{color}
\usepackage{tabularx}
\usepackage{epsfig}
\usepackage{graphicx}
\usepackage{wasysym}
\usepackage{dcolumn}
\usepackage{epstopdf}
\usepackage{subfigure}
\usepackage{latexsym}
\usepackage{psfrag}
\usepackage{cleveref}

\usepackage{float}

\usepackage{physics}
\usepackage{empheq,etoolbox}

\patchcmd{\subequations}
  {\theparentequation\alph{equation}}
  {\theparentequation.\alph{equation}}
  {}{}
\usepackage{graphicx,wrapfig}  
\begin{document}

\author{Johannes Gedeon}
\affiliation{Hannover Centre for Optical Technologies, Institute for Transport and Automation Technology (Faculty of Mechanical Engineering), and Cluster of Excellence PhoenixD,
Leibniz University Hannover, 30167 Hannover, Germany}
\author{Emadeldeen Hassan}
\affiliation{Department of Electronics and Electrical Communications, Menoufia University, Menouf 32952, Egypt}
\affiliation{Department of Applied Physics and Electronics, Umeå University, SE-901 87 Umeå, Sweden}
\author{Antonio Cal{\`a} Lesina}
\affiliation{Hannover Centre for Optical Technologies, Institute for Transport and Automation Technology (Faculty of Mechanical Engineering), and Cluster of Excellence PhoenixD,
Leibniz University Hannover, 30167 Hannover, Germany}

\title{Free-form inverse design of arbitrary dispersive materials in nanophotonics}

\date{\today}

\vspace{3cm}
\begin{abstract}

{\centering
\includegraphics[width=4cm]{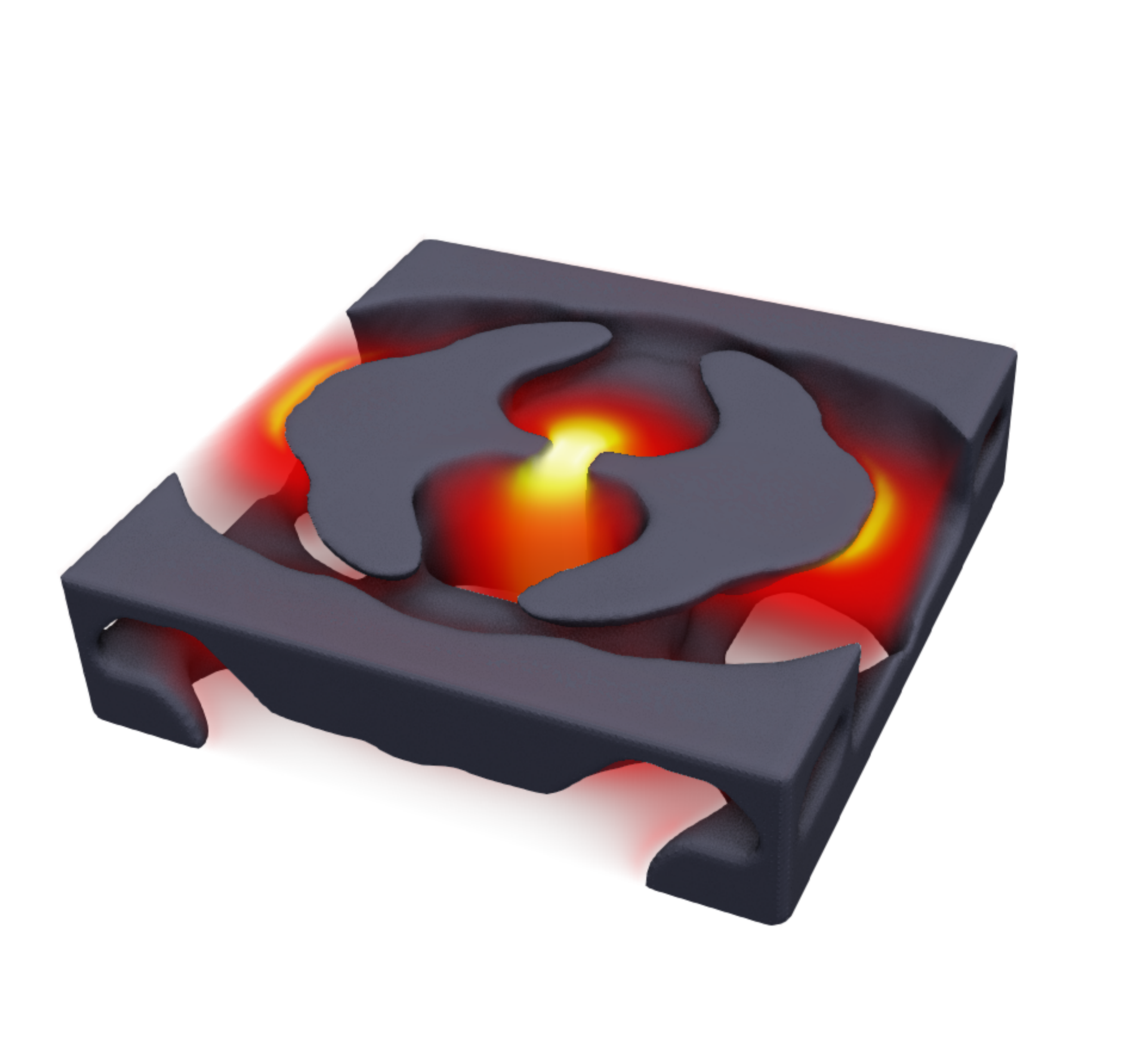}

}
In the last decades nanostructures have unlocked myriads of functionalities in nanophotonics by engineering light-matter interaction beyond what is possible with conventional bulk optics. The space of parameters available for design is practically unlimited due to the large variety of optical materials and geometries that can be realized by nanofabrication techniques. Thus, computational approaches are necessary to efficiently search for optimal solutions. In this paper, we enable the free-form inverse design in 3D of linear optical materials with arbitrary dispersion and anisotropy.
This is achieved by implementing the adjoint method based on the complex-conjugate pole-residue pair model within a parallel finite-difference time-domain solver, suitable for high-performance computing systems. 
Our method is tested on the canonical nanophotonic problem of field enhancement in a gap region. Also, our free-form designs of dispersive metallic and dielectric materials satisfy the fundamental curiosity of how optimized nanostructures look like in 3D. Unconventional free-form designs revealed by our method, although may be challenging or unfeasible with current technology, bring new insight into how light interacts with nanostructures, and could provide new ideas to inspire forward design.
\end{abstract}
\maketitle
\section{Introduction}\label{Sec:Introduction}
In the last decade, research in nanophotonics has enabled the manipulation and engineering of light-matter interaction at the nanoscale by means of nanostructured materials, such as metasurfaces and metamaterials. This was also made possible by the advances in nanofabrication technologies, such as two-photon polymerization, focused ion-beam milling, and lithographic methods, which have enabled the manufacturing of complex structures with features near or below the scale of the electromagnetic wavelength. These combined developments result in compact and efficient optical systems with new functionalities that are difficult or impossible to achieve using conventional bulk optical components, such as light structuring, beam steering, and dynamic optical control ~\cite{2018NaPho..12..659M}. The design space made available by fabrication techniques and materials is practically unlimited. Exploring large parameter spaces offers opportunities to find innovative designs with improved performance, or designs that can satisfy multi-objectives. However, exploring such large parameter spaces is computationally challenging. Numerical methods executed on computers can accelerate the design of optical systems beyond what is achievable via analytical methods and parametric sweeps, which usually start from an initial guess suggested by human intuition. In fact, parameters sweep or stochastic optimization methods, e.g., genetic algorithms, are only suitable to handle problems with few design parameters \cite{Non-grad}, and therefore only useful when a good initial guess is known.
In recent years, inverse design techniques have become popular in (nano-) photonics to automatically and efficiently explore large design spaces, and discover and optimize micro- and nanostructures with desired optical functionalities. Deep learning algorithms are emerging as a promising option for nanophotonics inverse design, but they require prohibitively large data sets for training \cite{DeepNNs1, DeepNNs2}. On the contrary, inverse design based on the adjoint method is more efficient since the gradient information used to update the design can be calculated with only two simulations \cite{AdjointBook}.\par 
Density-based topology optimization (TopOpt) for inverse design - originally introduced in mechanical engineering - is an iterative design process that allows us to optimize the distribution of a given material in a specified domain in order to optimize a certain objective function ~\cite{Bendsoe2004,TopOptForNanophotonics,Nomura07,Hassan14Topology,TopOptTutorial,Hassan20Multilayer}. 
This method has been applied to a variety of engineering problems in photonics, such as the optimization of metasurfaces to control certain properties of light (polarization, phase, angular momentum, achromatic focusing)~\cite{MetasurfacesReview, VortexBeams, Metalens1}, photonic crystals to find optimal omnidirectional band gaps \cite{PhotonicCrystal}, nanoantennas for broad-band enhancement \cite{Hassan:22}, small-scale particle accelerators \cite{LaserDriven}, quantum emitter and (de-) multiplexers as important components for photonic quantum computers~\cite{QE1, Beamsplitter, Demultiplexer, Multiplexer}, and non-linear photonic devices for e.g. second- and third-harmonic generation~\cite{NL2, NL1}.\par 
Topology optimization via the adjoint method has been presented in various forms based on frequency-domain formulations \cite{BookAdjoint, NonLinearTopOpt, PlasmonicTopOpt}. The treatment of dispersion by frequency-domain solvers only requires the use of a constant complex permittivity. However, we need to solve a linear system of equations for each frequency to obtain the response of a device, which entails that frequency domain methods are computationally less efficient when dealing with broadband performance. In addition, frequency-domain solvers typically have poor scalability for large-scale systems, thus limiting most designs to 2D. Hybrid time/frequency domain algorithms have recently been proposed, where FDTD was used to extract frequency-domain solutions, and the optimization routine itself is performed in the frequency domain \cite{meep, FDTDTopOpt}. For example, the method in \cite{meep} allows optimization for multiple frequencies using an epigraph formulation \cite{epigraf1, epigraf2}. This approach also has its drawbacks, as the computational cost increases with the number of frequencies for which the optimization algorithm is performed. 

The adjoint approach in the time domain is free from this limitation, since excitation by temporal signal is designed to contain all the frequencies of interest, which will ultimately result in broadband optimization \cite{Nomura07,Hassan2018,Hassan:22}. 
In addition, time-domain methods are in general more versatile for the optimization of time-dependent objectives, such as dynamic phenomena, pulse shaping, and non-stationary transient nonlinear effects. Moreover, time-domain methods scale nearly linearly on high-performance computing systems, thus enabling simulations that can handle optically large domains and/or highly refined mesh \cite{convergence}. Due to these scalability limitations, most of the optimizations performed in the frequency-domain focused on problems that can be decomposed by symmetry into 2D problems, or 3D designs consisting of planar sheets, with cylindrical symmetry, or with geometric invariance in one direction. Thus, the optimization of free-form nanostructures is still an open challenge \cite{Metasurfaces}.

\par
In this paper, we tackle two fundamental problems in topology optimization for nanophotonics: (1) the broadband inverse design of arbitrary dispersive materials, including anisotropy, and (2) the inverse design of free-form nanostructures in 3D. For the first point, we introduce a general adjoint scheme based on the time-dependent formulation of Maxwell’s equations and the complex-conjugate pole-residue pair (CCPR) model \cite{material}.
For the second point, we develop a fully parallel topology optimization algorithm by combining our parallel FDTD solver \cite{convergence} with a parallel MMA open-source routine \cite{MMA_Parallel}.\par
To date, most nanostructures for field enhancement are invariant in one direction. Although this is typically justified as a fabrication constraint, a move towards 3D free-form optimization is also missing due to the computational challenge associated with such 3D designs. 
However, fabrication technology is progressing, and it is timely to satisfy the fundamental curiosity in nanophotonics of what is the shape of optimized free-form nanostructures in 3D. In fact, nanostructures optimized for broadband response in a free-form fashion may exhibit shapes and geometries that have never been proposed before, which can later inspire traditional forward design. With this question in mind, we demonstrate the universality and efficiency of our method by optimizing free-form 3D metallic and dielectric nanostructures for broadband performance by choosing a canonical objective of maximizing the field enhancement in a gap region. Investigating the interaction of light with such computer-made complex designs can enable novel functionalities in nanophotonics, as the recently reported plasmonic anapole \cite{anapolePaper}.\par
In Section \ref{Sec:1} we discuss a model for arbitrary dispersion and anisotropy used within our adjoint formulation. In Section \ref{Sec:2}, we describe the optimization setup and the computational challenges associated to it. In Section \ref{Sec:3}, we present optimized free-form nanostructures for a selection of dispersive materials. Technical details about all derivations are finally provided in the Methods section \ref{Sec:Methods}.

\section{Inverse design of arbitrary dispersive materials }\label{Sec:1}
Most materials exhibit dispersion at optical frequencies, which enables various nanophotonic effects in plasmonics, epsilon-near-zero materials, and nonlinear optics. In the literature, various models are used to describe the materials dispersion, such as Debye, Drude, Lorentz, modified Lorentz, and Drude+Critical Points~\cite{material}.
These models can usually fit one type of material. 
The complex-conjugate pole-residue (CCPR)~model was recently proposed as a versatile model that can be used to fit any arbitrary material dispersion, including also the modelling of anisotropy~\cite{material}. 
We employ the CCPR model to inversely design dispersive materials aiming at a wideband performance of nanophotonic devices.

\begin{figure}[t!]
\includegraphics[width=1\linewidth]{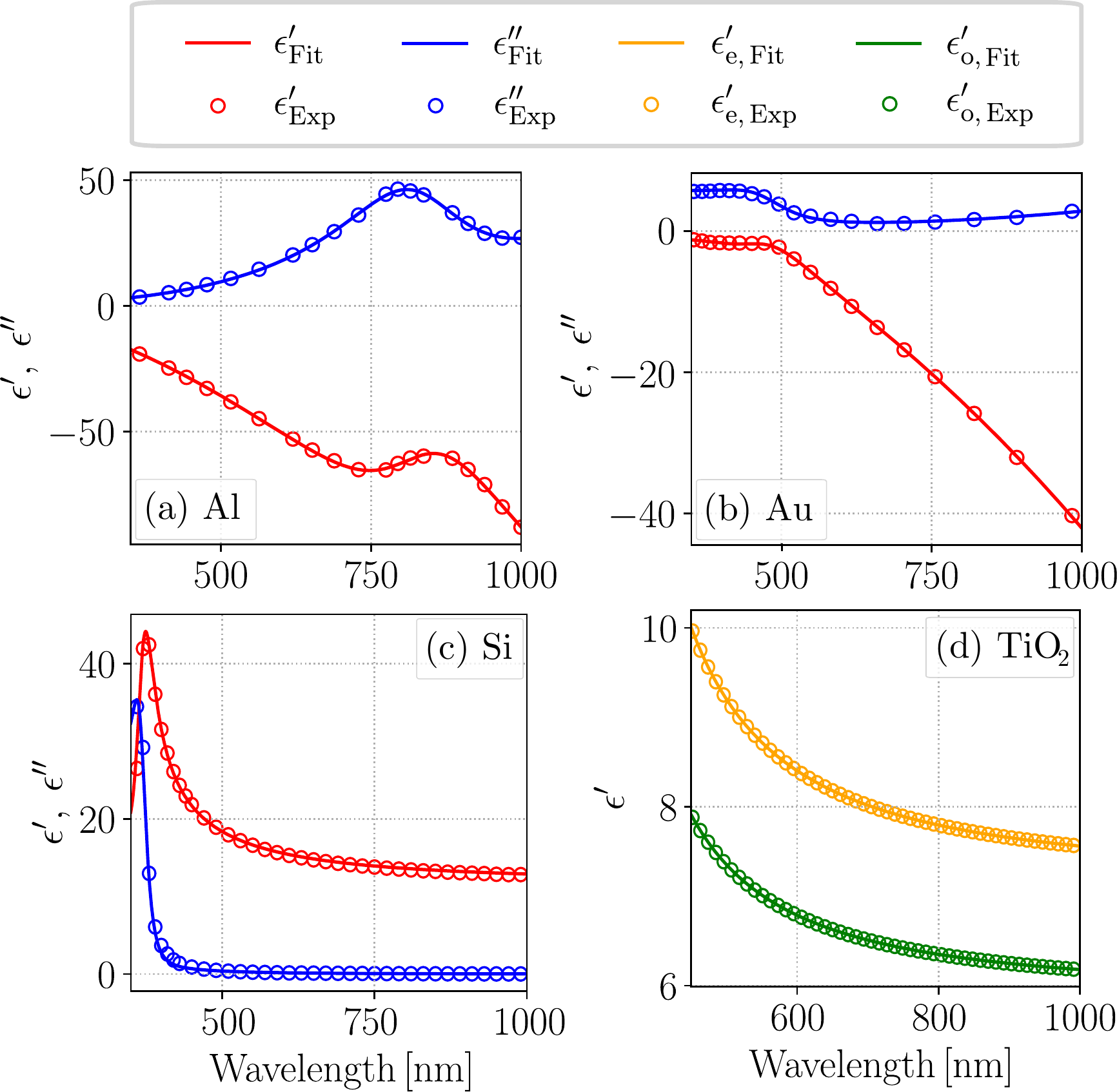}
\caption{\label{CCPR_Plot} Complex relative permittivity $\varepsilon'-j\varepsilon''$ of (a) Al, (b) Au, (c) Si and (d) anisotropic  TiO$_{\mathrm{2}}$ fitted using the CCPR model in expression~(\ref{Eq:CCPRModel}). We use the experimental reported in Refs.~\cite{Rakic:95, Au_Johnson, Si_Schinke, DeVore:51}, respectively. The corresponding CCPR parameters can be found in Table~\ref{Table}.}
\end{figure}

The complex relative permittivity tensor of the CCPR model for an anisotropic, dispersive medium is given by \cite{material}
\begin{equation}\label{Eq:CCPRModel}
\varepsilon_{\alpha \beta}(\omega)\!=\!\varepsilon_{\infty, \alpha \beta}+\frac{\sigma_{\alpha \beta}}{j \omega \varepsilon_0}+\sum_{p=1}^{P_{\alpha \beta}}\left(\frac{c_{p, \alpha \beta}}{j \omega-a_{p, \alpha \beta}}+\frac{c_{p, \alpha \beta}^*}{j \omega-a_{p, \alpha \beta}^*}\right)\!,
\end{equation}
where $\varepsilon_{\infty, \alpha \beta}$ is the relative permittivity at infinite frequency,  $\sigma_{\alpha \beta}$ is the static electric conductivity, and $\varepsilon_0$ is the vacuum permittivity. We assume $e^{\,j\omega t}$ time-dependency.
The indices $\alpha$ and $\beta$ denote the $x$, $y$ and $z$ component and $*$ represents the complex conjugation.
By a proper selection of its coefficients, the CCPR model can be used to incorporate all the standard dispersive models. 
Another strength of the model lies in its ability to accurately fit experimental permittivity data of materials using the vector fitting technique ~\cite{vectorFitting1, vectorFitting2} or
other related algorithms ~\cite{vectorFittingAlternative}.
Figures ~\ref{CCPR_Plot} (a)-(d) show the fitting of the experimental permittivity of Aluminium (Al) ~\cite{Rakic:95}, Gold (Au)~\cite{ Au_Johnson}, Silicon (Si)~\cite{Si_Schinke}, and  anisotropic Titanium Dioxide (TiO$_{\mathrm{2}}$)~\cite{DeVore:51}, respectively, using the CCPR model.
The spectral range 350-1000 nm corresponds to the range of interest for our broadband optimization. 
To fit the experimental data, we use three poles for Al and Au, two poles for Si, and a single pole for each axis of the anisotropic TiO$_{\mathrm{2}}$. The ordinary and extraordinary permittivities of Titanium Dioxide were fitted separately.
By choosing the number of the CCPR poles, we can compromise between the required fitting accuracy and the computational demand. The coefficients of the fitted model of the four materials are presented in Table~\ref{Table}.

\begin{table*}
 \centering      
{
\scriptsize \setlength{\tabcolsep}{2.7pt}
\begin{tabular}{ l  l  l  l  l l }\hline
     \textbf{Parameter}                   & Ag & Au                    & Si                    &      $\mathrm{TiO}_{2}\; (\varepsilon_{\mathrm{o}})$      & $\mathrm{TiO}_{2}\; (\varepsilon_{\mathrm{e}})$  \\\hline
 $\varepsilon_{\infty}$ &  \phantom{1.}3.07    &   \phantom{1.}2.31                     & \phantom{1.}1                         &  \phantom{1.}2.87                    & \phantom{1.}3.26  \\
 $\sigma$               &  $\phantom{-}1.49 \times 10^7$     &   $\phantom{-}1.21 \times 10^7$       & \phantom{1.}0                         & \phantom{1.}0                        & \phantom{1.}0  \\
 $a_1$                  &  $-1.89 \times 10^{14}$ & $-1.28 \times 10^{14}$           & $-8.00 \times 10^{14}+6.39 \times 10^{15} j$            & $-6.65 \times 10^{15} j$                   &  $-6.49 \times 10^{15} j$  \\     
 $c_1$                  &  $-1.00 \times 10^{18}$ & $-6.85 \times 10^{17} $          & $\phantom{-}7.31 \times 10^{14}-2.89 \times 10^{16} j$  & $\phantom{-}1.01 \times 10^{16} j$                   &  $\phantom{-}1.29 \times 10^{16} j$  \\   
 $a_2$                  &  $-5.46 \times 10^{14}-6.37 \times 10^{15} j$              & $-6.36 \times 10^{14}-3.89 \times 10^{15} j$            & $-2.32 \times 10^{14}+ 5.12 \times 10^{15} j$    &                  &  \\            
 $c_2$                  &  $\phantom{-}1.30 \times 10^{15}+1.54 \times 10^{15} j$    & $\phantom{-}2.06 \times 10^{15}+ 8.70 \times 10^{14} j$ & $\phantom{-}4.68 \times 10^{15}-4.55 \times 10^{15} j$     &                     &  \\
 $a_3$                  &  $-5.68 \times 10^{14}-3.43 \times 10^{14} j$              & $-2.96 \times 10^{15}-6.12 \times 10^{15} j$            &                       &                      &  \\
 $c_3$                  &  $\phantom{-}1.61 \times 10^{17}+1.00 \times 10^{12} j$    & $\phantom{-}1.60 \times 10^{13}+1.47 \times 10^{16} j$  &                       &                    &  \\\hline
\end{tabular}}
\caption{CCPR parameters for the permittivity spectral fitting in Fig.~\ref{CCPR_Plot}.}
\label{Table}
\end{table*}

In order to enable density-based topology optimization using the CCPR model, we need an interpolation strategy. In fact, 
the method requires the description of the material in the design domain as a spatial density distribution ($0\le\rho\le1)$ which is mapped to the material's permittivity, and consequently describes the topological shape of the photonic device as the density converges to a binary design. The interpolation scheme for the CCPR model is presented in the Methods section together with its incorporation into the time-domain Maxwell's equations. 
We emphasize that our developed topology optimization algorithm can target any arbitrary dispersion of linear materials, and it is only required to modify the model's coefficients and number of poles. In the next section we describe the optimization setup and the computational challenges associated to it, which results in the adoption of parallel computing in topology optimization. 

\section{Topology Optimization Setup}\label{Sec:2}

In order to handle the topology optimization of arbitrary dispersive materials, a considerable amount of computational resources is required in terms of memory. This is due to the fact that the electric field and the auxiliary fields associated with the CCPR model must be stored for all time steps in the entire design domain. The auxiliary fields are needed to model dispersion in the FDTD, and their number corresponds to the number of CCPR poles used to fit the material permittivity data. For example, storing the fields for the optimization of silicon nanostructures fitted with two CCPR poles, as shown in Fig.\,\ref{CCPR_Plot} (c), results in a three times memory consumption higher than the case of modeling a dispersionless dielectric material.

\par 
If we add to this our aim to design 3D free-form nanostructures, then the use of high-performance computing approaches becomes imperative.
Therefore, we implemented a fully parallel topology optimization algorithm within our parallel in-house FDTD solver with message passing interface (MPI) functionalities, whose nearly-linear scalability was tested up to 16k cores on a supercomputer \cite{convergence}. In our parallel software, the simulation region is divided into multiple subregions; each of them updates the fields at every voxel and exchanges the fields on its exterior surfaces with the adjacent subregions \cite{parallelism}. 
To incorporate topology optimization into our solver, we use the library developed by Niels Aage et al. \cite{petscTopOpt}.
This library builds on the PETSc\,ToolKit and provides an efficient implementation of filtering, projecting, and updating of the design using the method of moving asymptotes (MMA) in parallel \cite{MMA_Parallel}.

\begin{figure}[t!]
\includegraphics[width=1\linewidth]{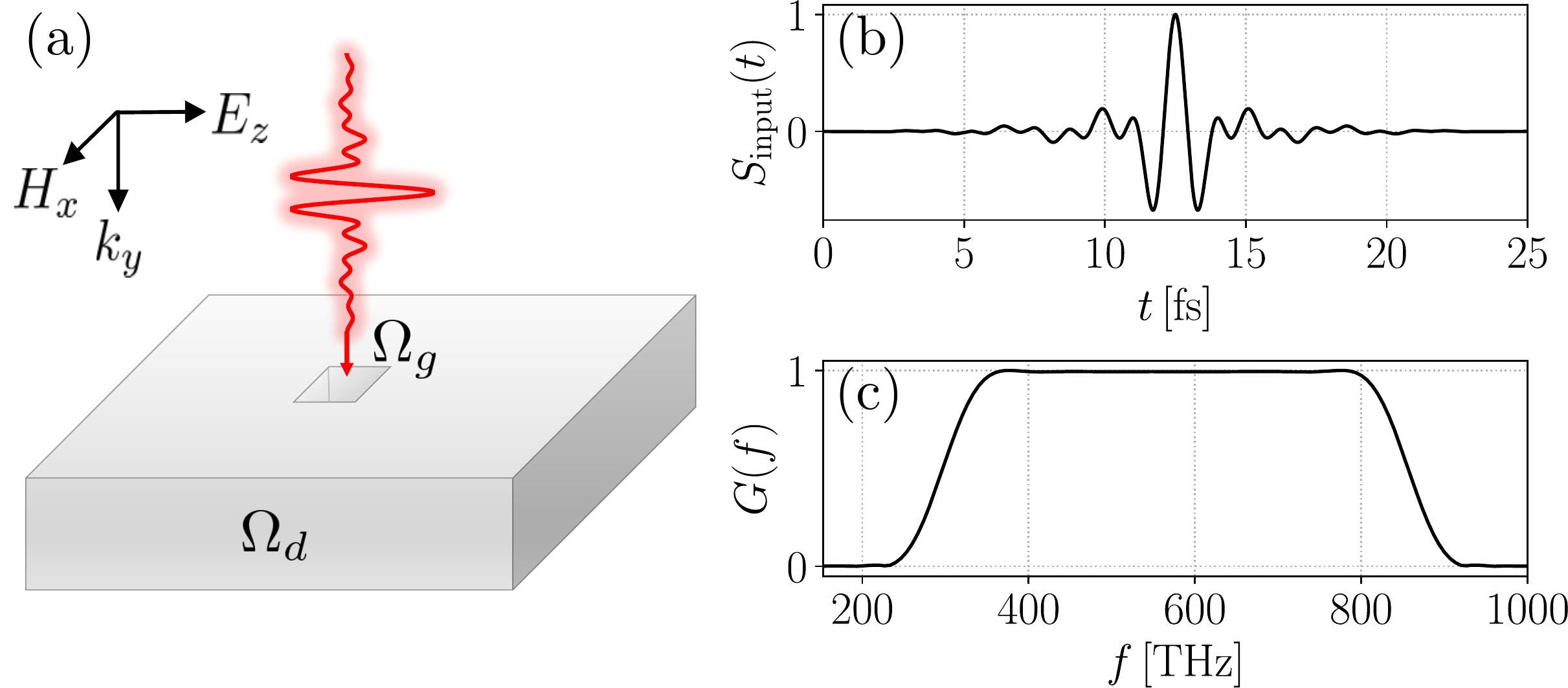}
\caption{\label{Signal} (a) Illustration of the optimization setup. The gap region $\Omega_{g}$ is located at the center of the design region $\Omega_{d}$. We excite the system using a $z$-polarized plane wave carrying a truncated sinc signal covering the spectral range of 350-1000\,nm. (b) Time and (c) frequency domain plots of the excitation source.}
\end{figure}

Here, we describe the optimization problem which leads to the 3D designs presented in the next section.
To demonstrate our method, we tackle a canonical problem in nanophotonics: the maximization of the field enhancement in the gap region of a nanostructure. This is motivated by the fundamental curiosity to uncover the shape of nanostructures optimized in 3D.
Fig.~\ref{Signal} (a) shows the optimization setup. 
We aim to maximize the electric field energy inside an observation volume $\Omega_{g}$ over a time duration $T$, by optimizing the topological structure of a given material in the design domain $\Omega_{d}$ that surrounds $\Omega_{g}$. 
During the forward simulation, the electric field energy in $\Omega_{g}$ is observed over the time duration $T$ until the fields are sufficiently decayed to ensure the convergence of the simulation.
The objective function, to be maximized, is defined as
\begin{equation}\label{Eq:Objective_analyitically_enhancement}
F[\mathbf{E}]:= \frac{\varepsilon_{0}}{2}\int_{\Omega_{g}}\int_{0}^{T} \varepsilon_{\infty, \mathrm{g}}\mathbf{E}\mathbf{E}\, \mathrm{d}t \mathrm{d}^{3}r,
\end{equation}
where we assume that the gap region contains a non-dispersive material, air in our case ($\varepsilon_{\infty, \mathrm{g}} = 1$). 

We fix the size of the design problem to $\Omega_{d}=100 \times 25 \times 100$ Yee cells for all investigated materials, which can provide a comparison in terms of the required computational demands. 
The observation region with a size of $\Omega_{g}=6 \times 25 \times 6$ Yee cells is located at the center of $\Omega_{d}$.
We chose a uniform spatial discretization in all directions with space step $\Delta_{\mathrm{D}}=5$\,nm for the optimization of Silicon and Titanium Dioxide, and $\Delta_{\mathrm{M}}=2$\,nm for Aluminium and Gold. 
This provides a physical domain size that is large enough to capture the different physical effects contributing to the local field enhancement, such as plasmonic and multipole resonances, and ensures enough accuracy for the simulation of all materials via the FDTD method~\cite{convergence}. 
All parameters used for the optimization as well as the simulation parameters based on the FDTD method are listed in Sec.\,\ref{Sec:FDTD}.\par

The excitation signal in Fig.~\ref{Signal} (b) is injected as a $z$-polarized plane wave, propagating along the $y$ direction. 
In time-domain topology optimization, the excitation signal defines the target bandwidth. As shown in Fig.~\ref{Signal} (b), the source consists of a sinc signal truncated to a few lobes with a bandwidth of $\sim$ 50 \% at half-maximum. Such signal modulates a carrier with a frequency of 576.5 THz that corresponds to the center of the spectral window of interest, and covers the spectral range 350-1000 nm, as shown in Fig.~\ref{Signal} (c). 
The excitation signal is also multiplied by a Hanning window to reduce the ripples in the excitation spectrum.

To enable gradient-based topology optimization, we use the adjoint-field method in the time domain to derive a gradient expression for our objective function, defined in expression\,\eqref{Eq:Objective_analyitically_enhancement}.
To compute the gradient of the objective function with respect to the material density, an additional adjoint simulation must be performed, which only differs from the forward simulation in the source of excitation. 
Both the forward and adjoint fields observed in the design region $\Omega_{d}$ are then used to compute the spatial gradient profile to update the design at each iteration step. 
The theoretical and technical details of our method are given in Section\,\ref{Sec:AdjFormulation} and the supplementary information.\par
In the next section, we demonstrate the flexibility of our inverse design method by aiming at free-form 3D designs enabled by our optimization algorithm for different optical materials with dispersion and anisotropy. 
We acknowledge that constraints such as length scale or structural invariance along a certain direction can be imposed into the algorithm. However, we chose not to impose any of such constraints to explore the free-form optimzation.


\section{Free-form 3D nanostructures}\label{Sec:3}

\begin{figure*}[t!]
\includegraphics[width=0.8\linewidth]{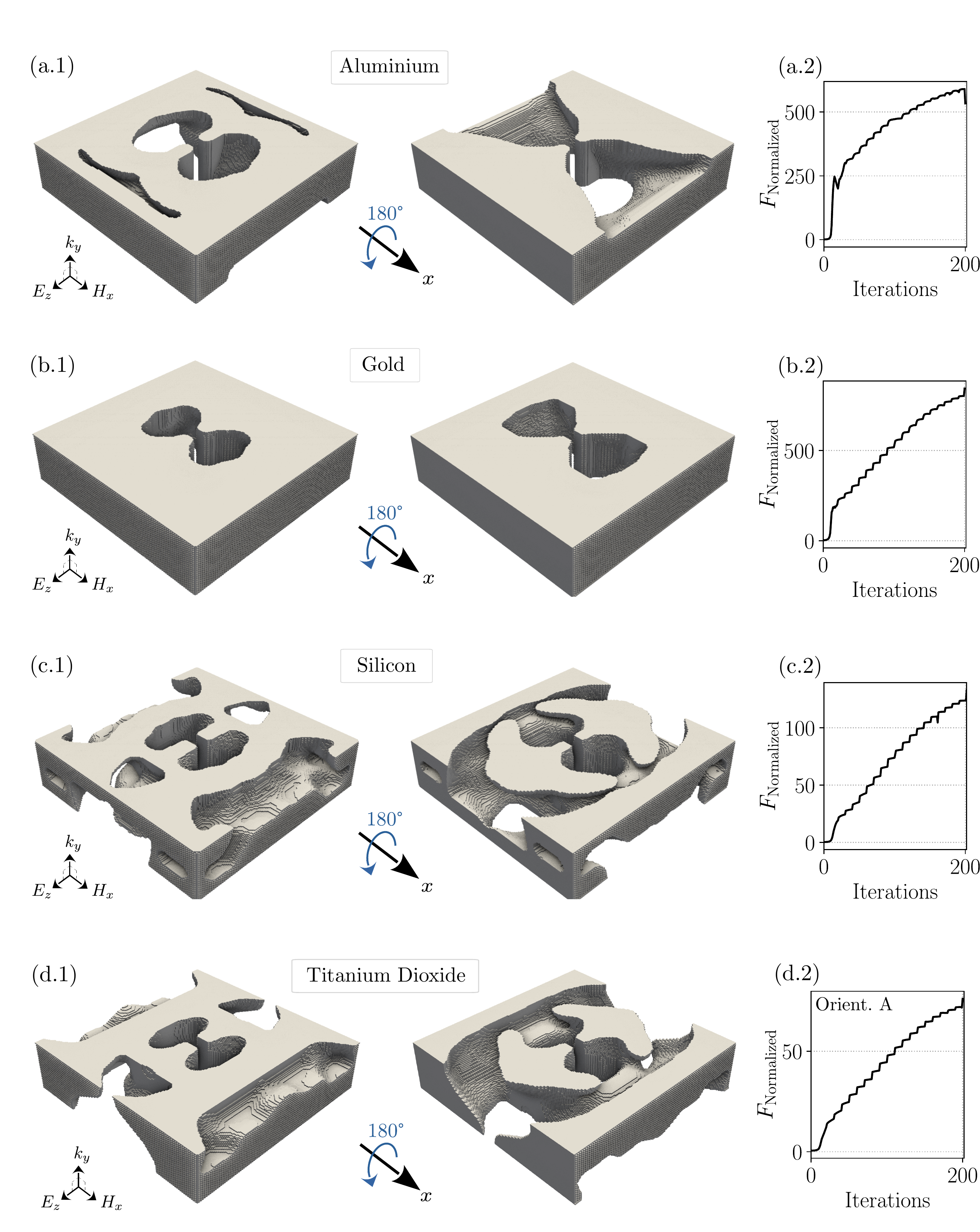}
\caption{\label{DesignPlot} (a.1) - (d.1) Top and bottom view of the topology optimized nanoantennas for Aluminium, Gold, Silicon and Titanium dioxide. 
(a.2) - (d.2) Corresponding progress of relative enhancement inside the gap versus the iteration number during the optimization. The nanoantennas are excited by a z-polarized broad band pulse propagating along the positive y-direction. The physical size of the metallic nanostructures is $\Omega_{d}=200\times50\times200\; \mathrm{nm}^3$ with a gap size of $\Omega_{g}=12\times50\times12\; \mathrm{nm}^3$. The size of the dielectric nanostructures is $\Omega_{d}=500 \times125\times500 \; \mathrm{nm}^3$ with a gap size of $\Omega_{g}=30\times125\times30\; \mathrm{nm}^3$. 
}
\end{figure*}

\begin{figure*}[t!]
\includegraphics[width=0.8\linewidth]{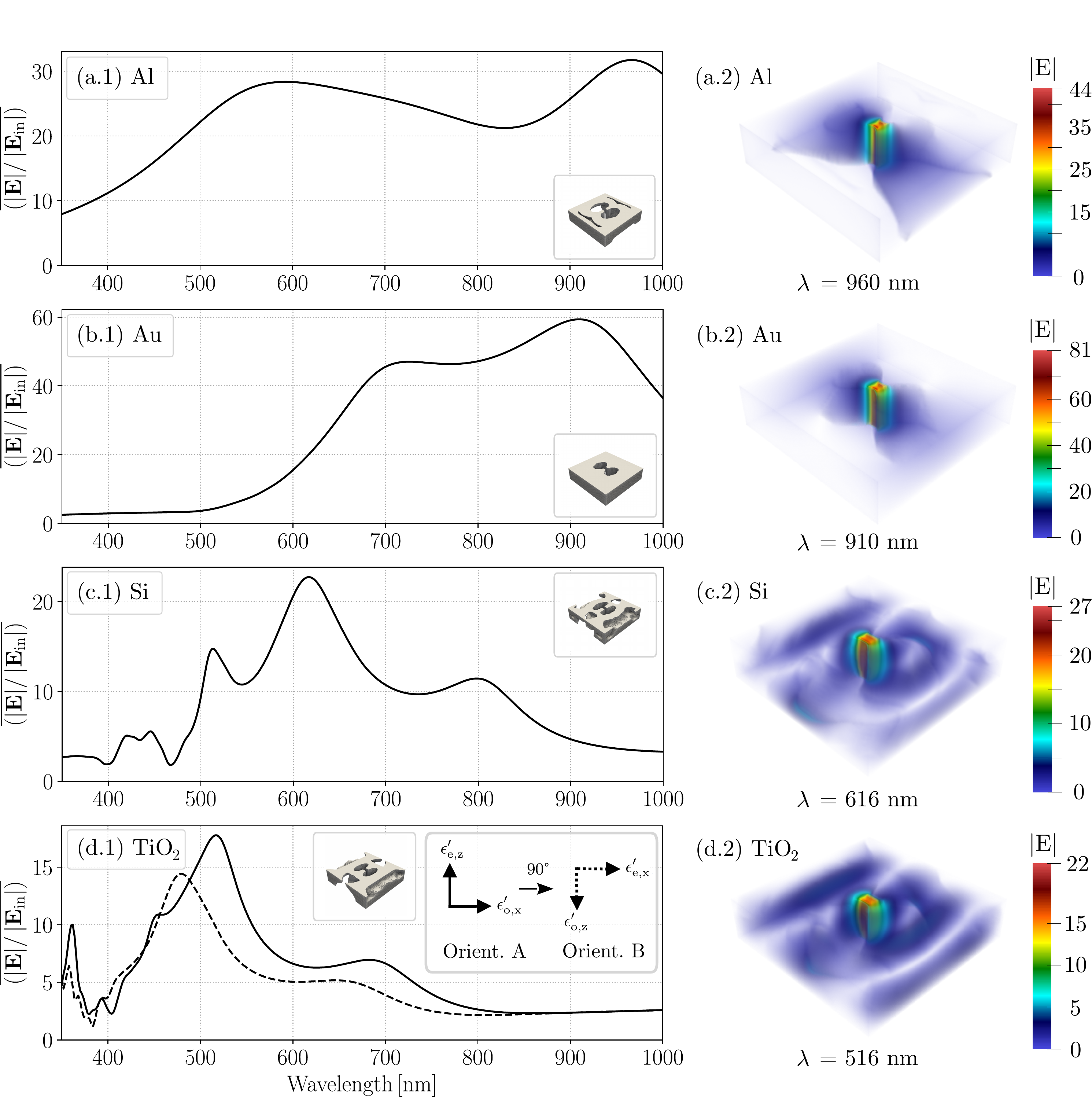}
\caption{\label{DFTPlot}Performance of the nanoantennas show in Fig.~\ref{DesignPlot}. (a.1) - (d.1): The averaged electric field enhancement inside the gap region over the spectral range of excitation. The dashed curve in (d.1) shows the performance of the TiO$_{2}$ antenna when the original orientation (Orient. A) of the anisotropy is rotated by 90° so that the lower ordinary dielectric constant $\epsilon^{\prime}_{\mathrm{o}}$ is aligned with the polarization of the incident excitation (Orient. B). (a.2) - (d.2): The corresponding field distribution for the wavelength for which the field enhancement has a maximum. }
\end{figure*}
\begin{figure*}[t!]
\includegraphics[width=0.75\linewidth]{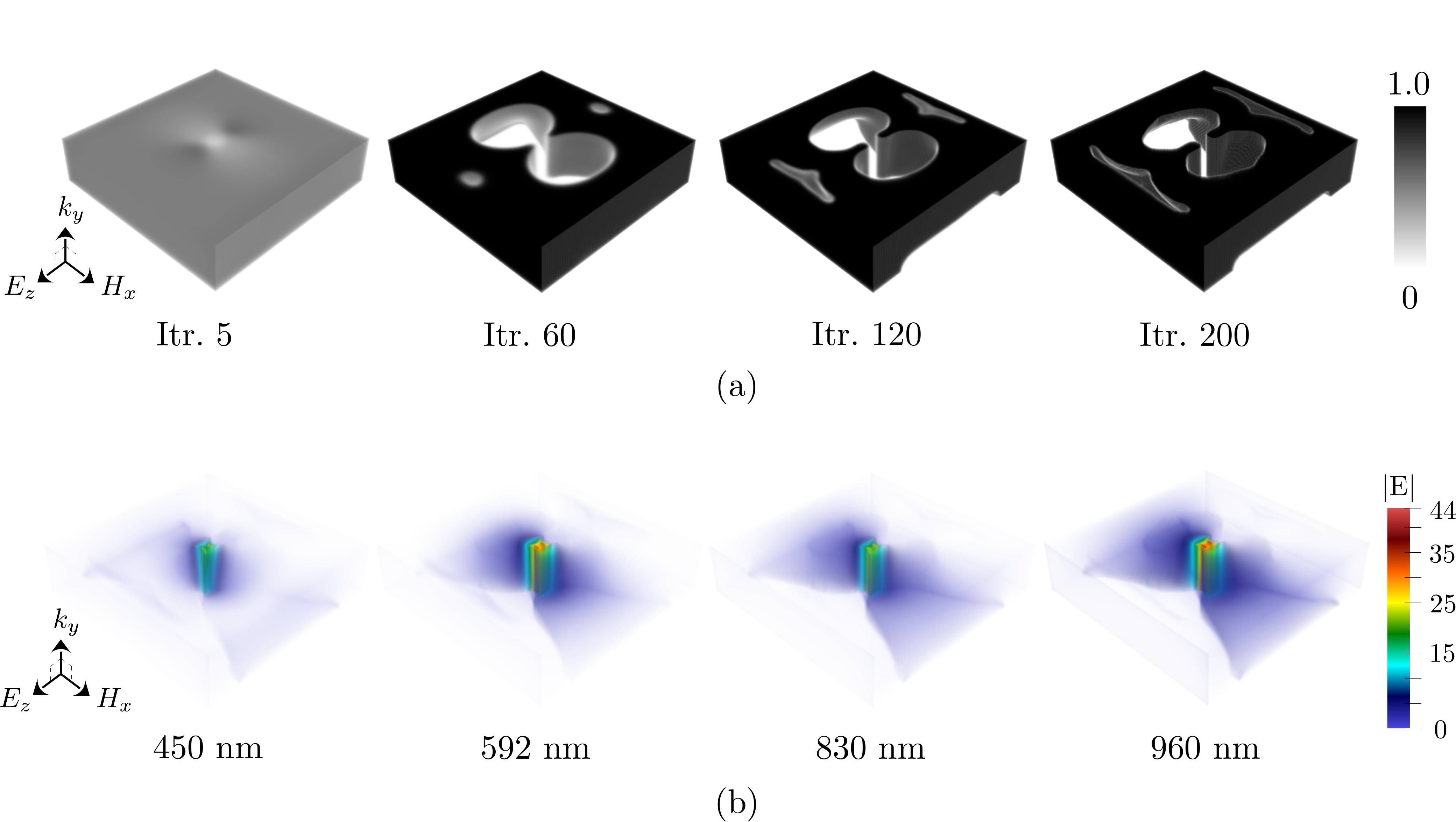}
\caption{\label{Al_DFT_and_Iterations} (a) Development of the filtered and projected density $\overline{\tilde{\rho}}$ 
during the optimization of the Aluminium nanoantenna presented in Fig.~\ref{DesignPlot} (a). The final binary design was obtained by thesholding $\overline{\tilde{\rho}}$ at iteration 200 with a projection value of $\eta=0.5$. (b) Spatial distribution of the electric field magnitude for different wavelengths within the optimized Aluminium nanoantenna shown in Fig.~\ref{DesignPlot} (a), including the maxima at 592 nm and 960 nm that appear in the frequency-domain response in Fig.~\ref{DFTPlot} (a).}
\end{figure*}
In this section, we present our free-form topology optimized dispersive metallic (Al and Au) and dielectric (Si and TiO$_{\mathrm{2}}$) nanonatennas for broadband field enhancement over the spectral range of 350 - 1000 nm. 
The designs obtained based on the optimization setup described in the previous section are illustrated in Fig.~\ref{DesignPlot}. All optimizations were performed for 200 iterations, to ensure a reasonable convergence and to enable a comparison of performance and computational requirements between different materials. 

We note the difference in shape between top and bottom views for the Al nanoantenna in Fig.~\ref{DesignPlot} (a.1), where the left subfigure shows typical features of plasmonic nanoantennas, such as a double-hole aperture, while the right subfigure shows features that are typical of metallic antennas in the microwave regime, such as a horn shape aperture. 
Also, the Au exhibits typical plasmonic geometrical features, such as a double-hole aperture, that is responsible for field localization and enhancement.
This shape was also recently reported to exhibit an anapole state \cite{anapolePaper}, which is also observed in this case (not shown). The dielectric nanostructures were optimized with a larger space step to allow, at parity of design domain size, more physical space for the nanostructure to develop. In general, dielectric nanostructures have a larger footprint compared to metallic ones at parity of resonant frequency. In the Si case, we observe the development of complex geometries with a profound difference between top and bottom views, where one side (left) shows an aperture similar to what observed for Au and the other side (right) shows a quasi free-standing nanoantenna.
The difference between top and bottom views is remarkable for all materials, and we think this is what enables the broadband performance. In the case of Au, although there is a difference in the size of the double-hole aperture in the top and bottom views, the shape of the double-hole does not change significantly, making the overall design very close to a lithographic-constrained structure.\par
As a measure of broadband performance, we computed the averaged electric field enhancement inside the gap region over the wavelength range 350 - 1000\,nm, as shown in Fig.~\ref{DFTPlot} (a.1) - (d.1). In addition, we captured the local field profile for the wavelengths for which the enhancement has a maximum in Fig.~\ref{DFTPlot} (a.2) - (d.2). The Gold and Aluminium nanoantennas yield a stronger enhancement than the Silicon and Titanium Dioxide nanoantennas, due to their plasmonic effects and the smaller gap size. From the field distribution plot, we see that the electric field barely penetrates into the bulk in the metallic cases compared to the dielectric designs. The Gold nanoantenna demonstrates the strongest enhancement, especially for high wavelengths. But in contrast to Aluminium, Gold shows a significantly poor performance at low wavelengths. This can be explained taking the physical properties of gold into account, such as the presence of absorption in this wavelength range, associated with the interband transition. In contrast, Aluminium is able to enhance the energy at low wavelengths more efficiently. We see a difference in the topological shape of both materials. The Aluminium developed more pronounced carving features compared to Gold to maximize the electric energy for the broadband pulse, and is apparently bounded by the physical size of the design domain. As a representative example, we plotted the development of the Aluminium design during the optimization and the field distributions for multiple wavelengths corresponding to the converged design (Fig.~\ref{Al_DFT_and_Iterations}).
Broadband field enhancement is achieved also in the case of the Si nanostructure, with several peaks due to multipole resonances. The field enhancement obtained for Si is lower due to the larger gap size (30 nm vs. 12 nm in Au and Al). Although both dielectric designs do not show such a significant difference in the topological shape, we acknowledge a weaker overall performance of Titanium Dioxide than Silicon. Comparing their materials properties in Figs.~\ref{CCPR_Plot} (c) and (d), we can attribute it to the fact that the real part of the permittivity of Silicon is higher than the ordinary and extraordinary values of Titanium Dioxide, enabling a more efficient local field enhancement. Also, Silicon shows weaker performance for $\lambda < 500$ nm, where it starts being absorptive. 
In addition, we studied the performance of the Titanium Dioxide nanoantenna, by rotating the anisotropic orientation - on which the antenna was originally designed - by 90° as shown in Fig.~\ref{DFTPlot} (d.1). The Titanium Dioxide nanoantenna was optimized based on an anisotropic axes orientation, where the extraordinary permittivity $\epsilon^{\prime}_{\mathrm{e}}$ axis is aligned with the $z$-polarized incident pulse, while the ordinary permittivity $\epsilon^{\prime}_{\mathrm{o}}$ axis is perpendicular to it and oriented along the $x$-axis. Since $\epsilon^{\prime}_{\mathrm{o}} < \epsilon^{\prime}_{\mathrm{e}}$, we obtain a shift of the spectral response to lower wavelengths. This effect is also well-known from Mie resonances and their appearance at certain wavelengths when changing the material's permittivity. 
The frequency-domain response is highly dependent on the underlying material and dispersion in the desired frequency range, as well as the physical size to which the design is constrained. We expect to see deviations in the frequency-domain response, when modifying the injected source in terms of spectral content and amplitude \cite{Hassan:22}.
This might enable better control over broadband enhancement, especially enforcing a better performance in regions of high absorption, but will consequently lead to a decrease of the stored electric energy (the objective) in the observation overall. 
One challenging aspect of time-domain optimization is the ability to precisely control the frequency-domain response of the device under test \cite{Hassan14Topology}. 
Attempts to use hybrid approaches to solve such challenging problems emphasize that this issue still requires further investigations~\cite{meep}.

\section{Methods}\label{Sec:Methods}
\subsection{Time-domain adjoint formulation}\label{Sec:AdjFormulation}
In this section, we formulate the density-based topology optimization problem for dispersive materials in time-domain.
We use the adjoint method to derive the gradient expression for an objective $F[\mathbf{E}]$, where $\mathbf{E}$ is the electric field that depends on a density distribution $\rho$ that represents our design variables. The objective function and its gradient are needed for gradient-based optimization algorithms. The full derivation is given in the supplementary material.

We formulate the conceptual optimization problem, 
\begin{equation}
\begin{aligned}
& \max _{\rho(\mathbf{r})} F[\mathbf{E}(\mathbf{r}, t)] \\
& \text { s.t. Maxwell's equations.}
\end{aligned}
\end{equation}
We assume non-magnetic materials. 
All the following derivations, however, can also be conducted for the magnetic permeability tensor and magnetic field $\mathbf{H}(\mathbf{r}, t)$.
The density is mapped to the physical material described by the CCPR model in expression~\eqref{Eq:CCPRModel}. 
In the following, we denote the design material with index $i=2$ and the background material with index $i=1$. For a given density value $\rho \in [0, 1]$, we apply a linear interpolation of the parameters and complex pole pairs in expression~\ref{Eq:CCPRModel} of the following form:
\begin{equation}\label{Eq:eps_mixed}
\varepsilon_{\infty, \alpha \beta}(\rho) := (1-\rho)\,\varepsilon_{\infty, \alpha \beta}^{(1)} +  \rho\, \varepsilon_{\infty, \alpha \beta}^{(2)},
\end{equation}
\begin{equation}\label{Eq:sigma_mixed}
\sigma_{\alpha \beta}(\rho) := (1-\rho)\,\sigma_{\alpha \beta}^{(1)} +  \rho\, \sigma_{\alpha \beta}^{(2)} + \rho\,(1-\rho)\,\gamma,
\end{equation}
\begin{equation}\label{Eq:Poles_mixed}
{\textstyle \sum_{{\alpha \beta}}}(\omega, \rho): = \sum_{i=1}^{2}\sum_{p=1}^{P_{\alpha \beta}^{(i)}}\kappa^{(i)}(\rho)\!\left(\frac{c_{p, \alpha \beta}^{(i)}}{j \omega-a_{p, \alpha \beta}^{(i)}}+\frac{c_{p, \alpha \beta}^{(i)*}}{j \omega-a_{p, \alpha \beta}^{(i)*}}\right)\!,
\medskip
\end{equation}
where $\kappa^{(1)}(\rho): = (1-\rho)$ and $\kappa^{(2)}(\rho): = \rho$. For the static conductivity in expression~(\ref{Eq:sigma_mixed}), we added an additional damping term $\gamma$, which can be tuned to avoid zero crossings of the permittivity (e.g. for metals) for intermediate values of $\rho$, inhibiting the optimization convergence \cite{Hassan:22, meep}.
The interpolated relative permittivity can be written as
\begin{equation}\label{Eq:Permitt_m gradients of the objective function with respect to thixed}
\varepsilon_{\alpha \beta}(\omega, \rho) = \varepsilon_{\infty, \alpha \beta}(\rho) + \sigma_{\alpha \beta}(\rho) + {\textstyle \sum_{{\alpha \beta}}}(\omega, \rho).
\end{equation}
For simplicity, we assume the permittivity of both design and background material to be diagonal,  $\varepsilon_{\alpha \beta} = 0$ for $\alpha \neq \beta$, and denote the spatial component of permittivity and fields by $k\in\{x,y,z\}$\par
For the formulation of the adjoint problem to maximize an objective $F[\mathbf{E}]$, we follow a similar approach as in Ref.~\cite{Hassan:22} with the same boundary conditions, but replacing the material model by our CCPR model. The electric field components $E_{k}$, $k \in \{x,y,z\}$, of the forward system driven by an incident pulse injected far away from the design and observation region can be obtained by solving the Maxwell's equations
\begin{widetext}
\begin{subequations}\label{Eq:Maxwells}
  \begin{empheq}[]{align}
(\nabla \times \mathbf{H})_{k} + \varepsilon_{0}\varepsilon_{\infty, k}(\rho)\partial_{t}E_{k} + \sigma(\rho)_{k} E_{k} + 2\sum_{i=1}^{2}\kappa^{(i)}(\rho)\Re\left\{\sum_{p=1}^{P_{k}^{(i)}}\partial_{t}Q_{p, k}^{(i)}\right\} &= 0,\\[1pt]
\text{For } i =1,2 \text{ and } \forall p \in 1, \ldots, P_k^{(i)}: \partial_{t}Q_{p, k}^{(i)}-a_{p, k}^{(i)}Q_{p, k}^{(i)}- c_{p, k}^{(i)}E_{k} &= 0, \\[8pt]
\mu_0 \partial_t H_k+(\nabla \times \mathbf{E})_k &=0,
\end{empheq}
\end{subequations}
\end{widetext}
where the complex auxiliary fields $Q_{p, k}^{(i)}$ must be computed for all poles $p \in {1,..., P_{k}^{(i)}}$ and for the corresponding component $k$ of both materials $i=1,2$. To compute the gradient of the  objective with respect to the density, an additional adjoint system must be solved, which differs from the forward system only in the source of excitation. The adjoint fields $\tilde{E}_k$, $\tilde{H}$ and $\tilde{Q}_{p, k}^{(i)}$ are obtained by solving the same set of equations, but introducing a source term $\mathbf{S}_{\mathrm{adj}}$ on the right-hand side of Eq.~(\ref{Eq:Maxwells}.a), which acts as the source for the adjoint system instead of the plane wave injected into the forward system. The adjoint source is the time-reversed (denoted by the symbol "$\overleftarrow{}$") functional derivative of the objective $F[\mathbf{E}]$ with respect to the forward field:
$\mathbf{S}_{\mathrm{adj}}:=\overleftarrow{\frac{\delta F[\mathbf{E}]}{\delta \mathbf{E}}}.$
The expression for the adjoint system as well as for the gradient can be obtained as follows (see the supplementary information). First,  we differentiate Eqs.~(\ref{Eq:Maxwells}) with respect to the density, and multiply Eq.~(\ref{Eq:Maxwells}a) by $\tilde{E_k}$, Eq.~(\ref{Eq:Maxwells}c) by $\tilde{H_k}$, and each of Eqs.~(\ref{Eq:Maxwells}b) by a corresponding term $-\frac{\kappa^{(i)}(\rho)}{\varepsilon_{0} c_{p, k}^{(i)}}\partial_{t}\tilde{Q}_{p, k}^{(i)}$, assuming a non-vanishing parameters $c_{p, k}^{(i)} \neq 0$. We further sum over all spatial components and integrate over time and space, taking the imposed boundary conditions of the forward fields into account. 
Then, we reverse the time $t \rightarrow \tau:= T-t$, and change the signs of the adjoint magnetic field and currents~\cite{Hassan15Time}, leads to the adjoint system as described before.
Moreover, we obtain the expression for the gradient of the objective function with respect to the density,
\begin{equation}\label{Eq:Gradient_analytic}
\begin{split}
\nabla_{\rho}F:=&\;-\int_{0}^{T}{ \sum_{k=1}^{3}}\varepsilon_{0}\partial_{\rho}\varepsilon_{\infty, k}(\rho) \overleftarrow{E_{k}}\partial_{t}\Tilde{E}_{k}\,\mathrm{d}t \\
 &-\int_{0}^{T} { \sum_{k=1}^{3}}\partial_{\rho}\sigma_{k}(\rho)\overleftarrow{E_{k}}\Tilde{E}_{k}\,\mathrm{d}t \\
 &+\int_{0}^{T}{ \sum_{k=1}^{3}}\sum^{2}_{i=1}\sum_{p=1}^{P_{k}^{(1)}}2\partial_{\rho}\kappa^{(i)}(\rho)\Re\left\{\partial_{t}\overleftarrow{Q}_{p, k}^{(i)}\right\}\Tilde{E}_{k}\,\mathrm{d}t.
\end{split}
\end{equation}
The derived expression also cover non-dispersive dielectric materials by simply setting all coefficients to 0, except for the $\varepsilon_{\infty,k}$ terms, leaving only the first term in expression~(\ref{Eq:Gradient_analytic}).
\subsection{FDTD implementation and design export}\label{Sec:FDTD}

\begin{figure}[t!]
\includegraphics[width=1.1\linewidth]{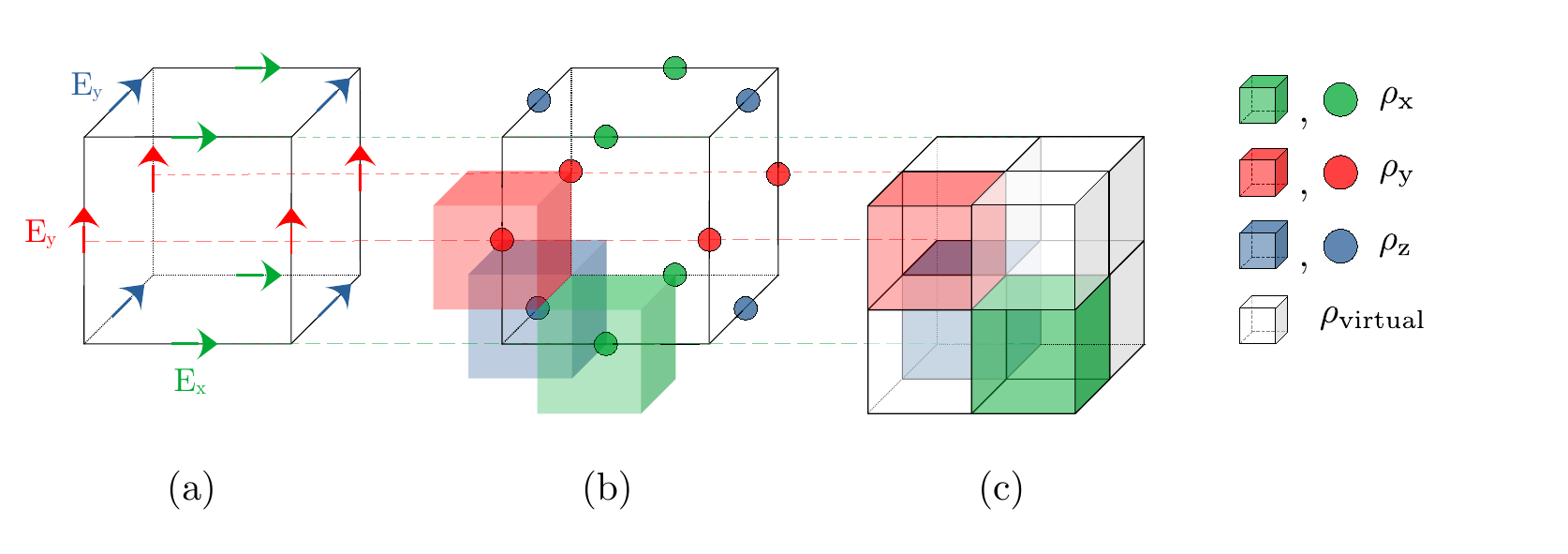}
\caption{\label{Interpolation} (a) According to the FDTD method, the electric field components are stored on a staggered grid. Each component $E_{\mathrm{k}}$ is placed at the center of the corresponding edge of the Yee cell. (b) Definition of subcells, being half as wide and locally offset by a quarter of the width of the Yee cell in each direction. Each of these cells encapsulate a density point $\rho_{\mathrm{k}}$ to the associated electric field component. (c) The empty space is filled with virtual subcells, encapsulating undefined density values $\rho_{\mathrm{virtual}}$. Their values must be assigned by a proper filtering, taking the defined values of neighboring subcells into account.}
\end{figure}

According to the FDTD method, each component of the electric and magnetic field is placed at different locations of the Yee cell (Fig.~\ref{Interpolation} (a)) \cite{Yee}. If the design region is divided into $N:= N_{x} \times N_{y}\times N_{z}$ Yee cells, we assign density values
$\rho_{k,n}, \; n\in \{1, ..., \tilde{N}_k\}$ at each location of the electric field components $E_k$, where $\tilde{N}_k$ is defined as
\begin{equation}
\tilde{N}_k:= \prod_{l\in\{x,y,z\}} (N_{l} + 1 - \delta_{kl}),  
\end{equation} 
with $\delta_{kl}$ denoting the Kronecker delta. Here, the additional $1 - \delta_{kl}$ cells ensure an accurate description of the design domain boundaries. As a consequence, we optimize $\tilde{N}:= \sum_{k\in\{x,y,z\}} \tilde{N}_k$ design variables in total.\newline\newline\newline\newline\newline
The update equations for the fields in Eqs.~(\ref{Eq:Maxwells}) within the FDTD framework are discussed in detail in Ref.~\cite{material}. The gradient of the objective function in expression~(\ref{Eq:Gradient_analytic}) with respect to each design variable $\rho_{k,n}$ can be discretized in a similar manner~\cite{Hassan15Time},
\begin{widetext}
\begin{equation}\label{Eq:Gradient_FDTD}
\begin{split}
\nabla_{\rho_{k,n}}F:=&-\varepsilon_{0}\left(\varepsilon_{\infty, k,n}^{(2)} - \varepsilon_{\infty, k,n}^{(1)}\right)\sum^{M}_{m=1}E^{M-m}_{k,n}\left(\tilde{E}^{m + \sfrac{1}{2}}_{k,n} - \tilde{E}^{m - \sfrac{1}{2}}_{k,n}\right) \\
& -\left(\sigma_{k,n}^{(2)} - \sigma_{k,n}^{(1)} + (1-2\rho_{k,n})\gamma\right)\sum^{M}_{m=1}E^{M-m}_{k,n}\frac{\tilde{E}^{m + \sfrac{1}{2}}_{k,n} + \tilde{E}^{m - \sfrac{1}{2}}_{k,n}}{2}\,\Delta t \\
& +2\,\sum^{M}_{m=1}\sum^{2}_{i=1}\sum_{p=1}^{P_{k}^{(i)}}\partial_{\rho}\kappa^{(i)}\Re\left\{Q^{M-m, (i)}_{p, k, n} - Q^{M-(m-1), (i)}_{p, k, n}\right\}\frac{\tilde{E}^{m + \sfrac{1}{2}}_{k,n} + \tilde{E}^{m - \sfrac{1}{2}}_{k,n}}{2},
\end{split}
\end{equation}
\end{widetext}

where $\partial_{\rho}\kappa^{(1)} = -1$ and $\partial_{\rho}\kappa^{(2)} = 1$, and $\Delta t$ denotes the time discretization. $M$ is the maximum number of time steps that corresponds to the simulation time. \par
When the optimization is complete and all density values are binarized after a proper thresholding, the design must be saved so that it is transferable to other (FDTD) software for simulation or to a software for planning the manufacturing, e.g. as STL file. To do so, we need to define a smallest cell around each density point so that their adjacent arrangement would fill the entire space without leaving any gaps or holes.

In this way, material or no material can be assigned to each cell to build and visualize the final design. Due to the lack of symmetry of the staggered Yee grid in 3D, the construction of such a cell is not possible. Therefore, we construct a symmetric material grid by dividing each Yee cell into 8 subcells, which are half as wide and locally offset by a quarter of the width of the original cell in each direction (Fig. \ref{Interpolation} (b) and (c)). 
In that way, each density point located at the edges of the Yee cell is encapsulated by a subcell. For the other 5 undefined, virtual subcells, we perform an interpolation, by averaging over the next neighbors, including the non-interpolated density points only. This interpolation scheme leaves the assignment of the original density points at the edges untouched and well-defined and is therefore transferable to any FDTD framework, assigning the material for the staggered grid in reverse. 

\subsection{Filtering and projection}\label{Sec:FilterAndProjection}
\begin{figure}[t!]
\includegraphics[width=1\linewidth]{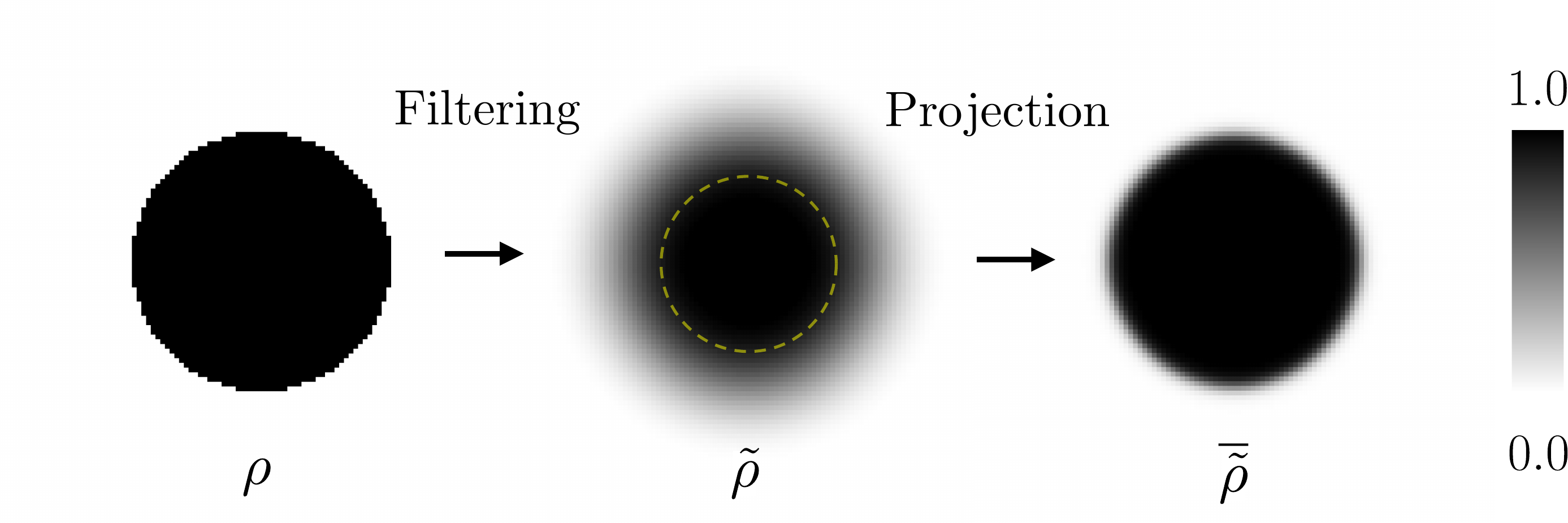}
\caption{\label{Filtering} Demonstration of the filtering and projection of the density in 2D. The original density $\rho$ representing a sphere with a radius $R_0$ is filtered according to Eq.~(\ref{FilterFunction}) with a filter radius of $R=0.3\,R_0$, marked as a yellow dashed circle. The filtered density $\tilde{\rho}$ will be projected using the smoothed Heaviside function in Eq.~(\ref{Projection}) to obtain $\overline{\tilde{\rho}}$, which is used as an input for the forward and adjoint simulation. Here, the projection parameters $\beta=10$ and $\eta=0.5$ were used.}
\end{figure}

Filtering is an effective way of introducing a weak sense
of length-scale into the design, to eradicate the appearance of single-pixel features \cite{BRUNS01Topology,Bourdin01Filters} or to cure the self-penalization issue when optimizing lossy structures~\cite{Hassan14Topology,Hassan2014patch,Aage17Topology}.
To control spatial design-field variations, we filter the design variables and apply a threshold procedure.  
Assuming we have $\tilde{N}$ design variables in total, we enumerate the variables with the index $n\in \{1,...,\tilde{N}\}$ in the following.  At each iteration step and for each density point $\rho_{n}$, we average over a neighborhood set of densities to obtain the filtered variable $\tilde{\rho}_n$,
\begin{equation}\label{FilterFunction}
\tilde{\rho}_n=\frac{\sum_{\rho_m \in \mathcal{B}^{n}_{R}} w\left(\mathbf{r}_n, \mathbf{r}_m\right)\rho_m}{\sum_{\rho_m  \in \mathcal{B}^{n}_{R}} w\left(\mathbf{r}_n, \mathbf{r}_m\right)},
\end{equation}
where $\mathcal{B}^{n}_{R}$ describes a sphere with radius $R$ around $\rho_n$. $w\left(\mathbf{r}_n, \mathbf{r}_m\right)$ is the weighting function defined as
\begin{equation}
w\left(\mathbf{r}_n, \mathbf{r}_m\right)=R-\left|\mathbf{r}_m-\mathbf{r}_n\right|.
\end{equation}
and ensures that values at a greater distance contribute less. Next, we project the filtered density using the smoothed Heaviside function
\begin{equation}\label{Projection}
\overline{\tilde{\rho}}_n=\frac{\tanh (\beta \eta)+\tanh \left(\beta\left(\tilde{\rho}_n-\eta\right)\right)}{\tanh (\beta \eta)+\tanh (\beta(1-\eta))}.
\end{equation}
The parameter $\eta$ determines the threshold value, and $\beta$ controls the sharpness of the projection, consequently leading  to binary design for $\beta \rightarrow \infty$. Fig.~\ref{Filtering} demonstrates the filtering and projecting procedure for a spherical density in 2D. 
The gradient of the objective with respect to the original density $\rho$ can be calculated using the chain,
\begin{equation}\label{GradientsFiltered}
\frac{\partial F}{\partial \rho_n}=\sum_{\rho_{m}\in \mathcal{B}^{n}_{R}} \frac{\partial F}{\partial \overline{\tilde{\rho}}_m} \frac{\partial \overline{\tilde{\rho}}_m}{\partial \widetilde{\rho}_m} \frac{\partial \tilde{\rho}_m}{\partial \rho_n}.
\end{equation}
The $\beta$ value will be increased during the optimization until the objective does not show any significant change, yielding an almost binary design. One way to tell whether an optimized design has converged to a discrete solution, is to use the measure of discreteness~\cite{Si07},
\begin{equation}\label{Eq:MeasureOfNon-discretness}
M_{\mathrm{n d}}=\frac{\sum_{n=1}^{\tilde{N}} 4 \,\overline{\tilde{\rho}}_n\left(1-\overline{\tilde{\rho}}_n\right)}{\tilde{N}} \times 100 \%.
\end{equation}
This measure is zero if the design only consists of elements with a 0 or 1 density, and gets maximized if all density points have an intermediate value of 0.5. 
Finally, the projected density is mapped to a binary design by thresholding with respect to the parameter $\eta$. The final design in the FDTD framework is exported by the interpolation scheme described in the previous section.\par

\subsection{Technical details of the optimization}\label{Sec:TopOptParams}
In this section, we list the parameters used for the topology optimization of the nanoantennas presented in the paper. It includes the simulation parameters for the FDTD method that were used to perform the forward and adjoint simulation, as well as the filter and projection variables to transform the density at each iteration step.\par
The design region $\Omega_{d}$ consists of $100 \times 25 \times 100$ Yee cells. The observation region with $6 \times 25 \times 6$ cells is located at the center of $\Omega_{d}$ (see Fig.~\ref{Signal} (a)). The design region is surrounded by convolutional perfectly matched layers (CPML) layers with a thickness of 20 Yee cells. We chose an isotropic spatial discretization $\Delta:=\mathrm{d}x=\mathrm{d}y=\mathrm{d}z$ and set its value small enough to ensure a sufficient accuracy while performing the FDTD simulations for all chosen materials \cite{convergence}. We set $\Delta_{\mathrm{D}}=5$ nm for the optimization of Silicon and Titanium Dioxide, and $\Delta_{\mathrm{M}}=2$ nm for Aluminium and Gold. 
We aim to optimize a $\Omega_{d}=200\times50\times200\; \mathrm{nm}^3$ Aluminium and Gold nanoantena with a gap size of $\Omega_{g}=12\times50\times12\; \mathrm{nm}^3$, and for a $\Omega_{d}=500 \times125\times500 \; \mathrm{nm}^3$ Silicon and Titanium Dioxide nanonatenna with a gap size of $\Omega_{g}=30\times125\times30\; \mathrm{nm}^3$. 
We chose a time discretization of $\mathrm{d}t=8.34\times10^{-3}$ fs for both dielectrics and performed forward and adjoint simulations for 10000 time steps, which corresponds to a time duration $T=83.40$ fs in Eq.~(\ref{Eq:Objective_analyitically_enhancement}). For metals, we chose $\mathrm{d}t=3.34\times10^{-3}$ fs and 18000 time steps, corresponding to $T=60.12$ fs. This parameter setting ensured Courant stability and a sufficient decay of the fields so that the objective and gradient did not show significant change at the end of each forward or adjoint simulation.\par
For all materials, an artificial damping $\gamma=50000$ was chosen in Eq.~(\ref{Eq:sigma_mixed}), to ensure convergence. The proper choice of this parameter is discussed in Ref.~\cite{Hassan:22}. For the radius of the filter, we chose $R=12$ Yee cells, corresponding to a radius of $R_{\mathrm{D}}=60$ nm for dielectrics and $R_{\mathrm{M}}=24$ nm for metals. Starting with an initial projection value $\beta_0=10^{-4}$, we run each optimization for 200 iterations, increasing $\beta$ up to a value of $\beta_{\mathrm{max}}=52$. The thresholding value was set to $\eta=0.5$. The maximum number of iteration together with $\beta_{\mathrm{max}}$ ensured, that the density is $\overline{\tilde{\rho}}$ does not contain larger gray areas of intermediate density values between 0 and 1 before thresholding, i.e. performing a binarization of $\overline{\tilde{\rho}}$. For verification, we computed the measure of discreteness of the final designs according to Eq.~(\ref{Eq:MeasureOfNon-discretness}), and obtained: $M^{\mathrm{Al}}_{\mathrm{n d}}=1.0229\,\%$, $M^{\mathrm{Au}}_{\mathrm{n d}}=0.3918\,\%$, $M^{\mathrm{Si}}_{\mathrm{n d}}=1.8153\,\%$, and $M^{\mathrm{TiO_2}}_{\mathrm{n d}}=1.1907\,\%$.
The densities were updated by the method of moving asymptots (MMA) \cite{Svanberg1987}.\par
The optimizations were performed with our in-house FDTD code, and we use the commercial software \textit{Ansys Lumerical} to cross verify the performance of the final optimized designs. 
The simulations have been performed on the supercomputer \textit{HLRN-IV-System Emmy} in G\"ottingen, Germany, provided by the \textit{North German Supercomputing Alliance} as part of the National High Performance Computing (NHR) infrastructure. 
It took $\approx$ 2-14 hours to optimize the structures presented in this paper, using 1536 cores (Cascade 9242, HLRN-IV-System Emmy in G\"ottingen, Germany). 
The computation time depends on the number of the CCPR poles used in the simulation.

\section{Conclusion}\label{Sec:Conclusions}
We introduced a general density-based topology optimization approach to design arbitrary dispersive and anisotropic nanophotonic designs.
The optimization problem is formulated in the time domain based on the CCPR model.
We employ the auxiliary equations approach to Maxwell's equations and derive an adjoint system that allows to compute the gradient by using only two system solutions.
By choosing the number of CCPR poles, a compromise between the required fitting accuracy and the computational requirements can be found. The method was implemented in our highly paralleized  FDTD code, and we provide an interpolation scheme to extract the final design from the staggered Yee grid. 
The reliability of the method is verified by designing dielectric and metallic nanoantennas for broadband enhancement, enabling free-form optimization in 3D.
By combining parallel topology optimization and parallel FDTD solver, we unlock not only the design of arbitrary dispersive materials, but also the free-form optimization of nanostructures.
Despite the computational effort required for free-form optimization, and the limitations of current lithographic techniques to fabricate such designs, nanofabrication in 3D is advancing, for example via additive manufacturing and two-photon polymerization.
Thus, it is timely and inspiring to explore how optimized free-form nanostructures look like in 3D.
Time-domain based optimization has been less studied and applied than frequency-dependent methods.
Our contribution holds a great potential not only for optimizing dispersive nanostructures and broadband response, but also for a variety of other design problems where time-dependent objectives are appropriate, such as dynamic phenomena, transient nonlinear effects, or even time-varying materials.

\section{Competing interests} 

The authors declare no competing interests.

\section{Acknowledgments} 
We acknowledge the computing time granted by the Resource Allocation Board and provided on the supercomputer Lise and Emmy at NHR@ZIB and NHR@Göttingen as part of the NHR infrastructure. The calculations for this research were conducted with computing resources under the project nip00059. We acknowledge the central computing cluster operated by Leibniz University IT Services (LUIS) at the Leibniz University Hannover. We acknowledge the Deutsche Forschungsgemeinschaft (DFG, German Research Foundation) under Germany’s Excellence Strategy within the Cluster of Excellence PhoenixD (EXC 2122, Project ID 390833453). A.C.L. acknowledges the German Federal Ministry of Education and Research (BMBF) under the Tenure-Track Program. The publication of this article was funded by the Open Access Fund of the Leibniz Universität Hannover.


\bibliography{main.bib}

\end{document}


\author{Johannes Gedeon}
\affiliation{Hannover Centre for Optical Technologies, Institute for Transport and Automation Technology (Faculty of Mechanical Engineering), and Cluster of Excellence PhoenixD,
Leibniz University Hannover, 30167 Hannover, Germany}
\author{Emadeldeen Hassan}
\affiliation{Department of Electronics and Electrical Communications, Menoufia University, Menouf 32952, Egypt}
\affiliation{Department of Applied Physics and Electronics, Umeå University, SE-901 87 Umeå, Sweden}
\author{Antonio Cal{\`a} Lesina}
\affiliation{Hannover Centre for Optical Technologies, Institute for Transport and Automation Technology (Faculty of Mechanical Engineering), and Cluster of Excellence PhoenixD,
Leibniz University Hannover, 30167 Hannover, Germany}

\title{Supplementary information for \\
``Free-form inverse design of arbitrary dispersive materials in nanophotonics''}

\date{\today}
\maketitle

\section{Derivation of the adjoint equations} 
Here, we derive the adjoint equations for density-based topology optimization based on the time domain Maxwell equations and the Complex-conjugate Pole-residue material model with $\exp^{\,j\omega}$ time-dependency
\begin{equation}
\varepsilon_{\alpha \beta}(\omega)=\varepsilon_{\infty, \alpha \beta}+\frac{\sigma_{\alpha \beta}}{j \omega \varepsilon_0}+\sum_{p=1}^{P_{\alpha \beta}}\left(\frac{c_{p, \alpha \beta}}{j \omega-a_{p, \alpha \beta}}+\frac{c_{p, \alpha \beta}^*}{j \omega-a_{p, \alpha \beta}^*}\right),
\end{equation}
where $\varepsilon_{\infty, \alpha \beta}$ is the relative permittivity at infinite frequency, and $\sigma_{\alpha \beta}$ is the static conductivity. The indices $\alpha$ and $\beta$ denote the $x$, $y$ and $z$ component and $*$ represents the complex conjugation. In the following, we denote the design material with index $i=2$ and the background material with index $i=1$. For a given density value $\rho \in [0, 1]$ we apply a linear interpolation of the parameters and complex pole pairs of the following form:
\begin{subequations}\label{Eq:InterpolatedModel}
  \begin{empheq}[]{align}
\varepsilon_{\infty, \alpha \beta}(\rho) &:= (1-\rho)\,\varepsilon_{\infty, \alpha \beta}^{(1)} +  \rho\, \varepsilon_{\infty, \alpha \beta}^{(2)},\\
\sigma_{\alpha \beta}(\rho) &:= (1-\rho)\,\sigma_{\alpha \beta}^{(1)} +  \rho\, \sigma_{\alpha \beta}^{(2)} + \rho\,(1-\rho)\,\gamma,\\
{\textstyle \sum_{{\alpha \beta}}}(\omega, \rho) &:= \sum_{i=1}^{2}\kappa^{(i)}(\rho)\sum_{p=1}^{P_{\alpha \beta}^{(i)}}\left(\frac{c_{p, \alpha \beta}^{(i)}}{j \omega-a_{p, \alpha \beta}^{(i)}}+\frac{c_{p, \alpha \beta}^{(i)*}}{j \omega-a_{p, \alpha \beta}^{(i)*}}\right),
\end{empheq}
\end{subequations}
where $\kappa^{(1)}(\rho): = (1-\rho)$ and $\kappa^{(2)}(\rho): = \rho$. According to these equations, the interpolated relative permittivity can be written as
\begin{equation}
\varepsilon_{\alpha \beta}(\omega, \rho) = \varepsilon_{\infty, \alpha \beta}(\rho) + \frac{\sigma_{\alpha \beta}(\rho)}{j \omega \varepsilon_0} + {\textstyle \sum_{{\alpha \beta}}}(\omega, \rho).
\end{equation}
We assume that the \textit{figure of merit} (or \textit{objective}) we aim to maximize, is a functional of the electric field only, such that the optimization problem can be formulated as follows:
\begin{equation}
\begin{aligned}
& \max _{\rho} F[\mathbf{E}] \\
& \text { s.t. Maxwell's equations,}
\end{aligned}
\end{equation}
where boundary conditions and manufacturability constraints can be included. The functional derivate of the objective $F[\mathbf{E}]$ with respect to the density is
\begin{equation}\label{Eq:ObjectiveDerivative}
\frac{\delta F[\mathbf{E}]}{\delta \rho}=\frac{\delta F[\mathbf{E}]}{\delta \mathbf{E}} \cdot \frac{\mathrm{d} \mathbf{E}}{\mathrm{d} \rho},
\end{equation}
where the first multiplicator on the right hand side denotes the functional derivate of the objective function with respect to the electric field. We denote the derivative of any local function with respect to the density by ``$\mathrm{d}_{\rho}$'' in the following.\par 
We assume a diagonal permittivity tensor and non-magnetic materials in the following. 
We denote the spatial physical domain with $\Omega$, and the time interval as $I=[0,T]$, and consider an excitation of the \textit{forward} system by a pulse injected at $\partial \Omega$ at $t=0$, and vanishing fields for $t\in \partial I$. The full system of Maxwell equations for each spatial component $k\in\{1,2,3\}$ and $\forall (\mathbf{r},t) \in \Omega \times I$ of the forward system reads \cite{material}:
\begin{subequations}\label{Eq:Maxwells2}
  \begin{empheq}[]{align}
-(\nabla \times \mathbf{H})_{k} + \varepsilon_{0}\varepsilon_{\infty, k}\partial_{t}E_{k} + \sigma_{k} E_{k} + 2\sum_{i=1}^{2}\sum_{p=1}^{P_{k}^{(i)}}\kappa^{(i)}\Re\left\{\partial_{t}Q_{p, k}^{(i)}\right\} &= 0,\\[1pt]
\text{For } i =1,2 \text{ and } \forall p \in 1, \ldots, P_k^{(i)}: \partial_{t}Q_{p, k}^{(i)}-a_{p, k}^{(i)}Q_{p, k}^{(i)}- \varepsilon_{0}c_{p, k}^{(i)}E_{k} &= 0.\\[8pt]
\mu_0 \partial_t H_k+(\nabla \times \mathbf{E})_k &=0.
\end{empheq}
\end{subequations}
We emphasize our chosen material interpolation in Eq.~(\ref{Eq:InterpolatedModel}), the parameters $a_{p, k}^{(i)}$ and $c_{p, k}^{(i)}$ do \textit{not} depend on the density itself. In contrast, all fields depend implicitly on $\rho$, and the local functions interpolating the material such as $\varepsilon_{\infty, k}$, $\sigma_{k}$ and $\kappa^{(i)}$ depend directly on the spatial density distribution.\par
We define adjoint fields $\mathbf{\tilde{E}}$, $\mathbf{\tilde{H}}$ and $\tilde{Q}_{p, k}^{(i)}$, for $i =1,2$, $k\in\{1,2,3\}$ and $\forall p \in 1, \ldots, P_k^{(i)}$,
sharing the same properties as the forward fields, i.e. the electric and magnetic adjoint fields are real, and the adjoint auxiliary fields are allowed to be complex. 
We derivate the system of Eqs.~(\ref{Eq:Maxwells2}) with respect to $\rho$, and multiply Eq.~(\ref{Eq:Maxwells2}a) by $\tilde{E_k}$, Eq.~(\ref{Eq:Maxwells2}c) by $\tilde{H_k}$, and each of Eqs.~(\ref{Eq:Maxwells2}b) by a corresponding term $-\kappa^{(i)}\frac{\partial_{t}\tilde{Q}_{p, k}^{(i)}}{\varepsilon_{0} c_{p, k}^{(i)}}$, assuming a non-vanishing parameters $c_{p, k}^{(i)} \neq 0$. Furthermore, we sum over the spatial components and get
\begin{equation}\label{Eq:MaxwellsDerivE}
\begin{split}
-\sum_{k=1}^{3}\left\{(\nabla \times \mathrm{d}_{\rho}\mathbf{H})_{k}\tilde{E}_k 
+ \varepsilon_{0}(\mathrm{d}_{\rho}\varepsilon_{\infty, k})\tilde{E}_k \partial_{t}E_{k}
+ \varepsilon_{0}\varepsilon_{\infty, k}\tilde{E}_k \partial_{t}(\mathrm{d}_{\rho}E_{k})\right\} \;&\\[8pt]
+ \sum_{k=1}^{3}\left\{(\mathrm{d}_{\rho}\sigma_{k}) \tilde{E}_k E_{k}
+ \sigma_{k} \tilde{E}_k (\mathrm{d}_{\rho}E_{k})\right\}\;&\\[8pt]
+ \sum_{k=1}^{3}\sum_{i=1}^{2}\sum_{p=1}^{P_{k}^{(i)}}2(\mathrm{d}_{\rho}\kappa^{(i)})\tilde{E}_{k}\Re\left\{\partial_{t}Q_{p, k}^{(i)}\right\}  
+ \sum_{k=1}^{3}\sum_{i=1}^{2}\sum_{p=1}^{P_{k}^{(i)}}2\kappa^{(i)}\tilde{E}_{k}\Re\left\{\partial_{t}(\mathrm{d}_{\rho}Q_{p, k}^{(i)})\right\} &= 0,
\end{split}
\end{equation}
\begin{equation}\label{Eq:MaxwellsDerivH}
\begin{split}
\phantom{\;xxxxxxxxxxxxxxxxxxxxxxxxxxxxxxx}\sum_{k=1}^{3}\left\{\mu_0 \tilde{H_k}\partial_t \mathrm{d}_{\rho}H_k+ \tilde{H_k}(\nabla \times \mathrm{d}_{\rho}\mathbf{E})_k\right\} &=0.\\
\end{split}
\end{equation}
And for $i =1,2$ and $\forall p \in 1, \ldots, P_k^{(i)}$ we obtain:
\begin{equation}\label{Eq:MaxwellsDerivQ}
\begin{split}
\sum_{k=1}^{3}\left\{\frac{-\kappa^{(i)}}{\varepsilon_{0} c_{p, k}^{(i)}}\partial_{t}\tilde{Q}_{p, k}^{(i)}\partial_{t}(\mathrm{d}_{\rho}Q_{p, k}^{(i)})
+\frac{\kappa^{(i)} a_{p, k}^{(i)}}{\varepsilon_{0} c_{p, k}^{(i)}}\partial_{t}\tilde{Q}_{p, k}^{(i)} (\mathrm{d}_{\rho}Q_{p, k}^{(i)})
+\kappa^{(i)}\partial_{t}\tilde{Q}_{p, k}^{(i)}(\mathrm{d}_{\rho}E_{k})\right\} &= 0.
\end{split}
\end{equation}
By addition of the the complex conjugates (denoted as ``c.c.'') of Eqs.~(\ref{Eq:MaxwellsDerivQ}), and summing over the indices $i$ and $p$, we reduce the equations above to:
\begin{equation}\label{Eq:Qsum}
\begin{split}
\sum_{k=1}^{3}\sum_{i=1}^{2}\sum_{p=1}^{P_{k}^{(i)}}\left\{\frac{-\kappa^{(i)}}{\varepsilon_{0} c_{p, k}^{(i)}}\partial_{t}\tilde{Q}_{p, k}^{(i)}\partial_{t}(\mathrm{d}_{\rho}Q_{p, k}^{(i)})
+\frac{\kappa^{(i)} a_{p, k}^{(i)}}{\varepsilon_{0} c_{p, k}^{(i)}}\partial_{t}\tilde{Q}_{p, k}^{(i)} (\mathrm{d}_{\rho}Q_{p, k}^{(i)})+\mathrm{c.c.}\right\} \\
+\sum_{k=1}^{3}\sum_{i=1}^{2}\sum_{p=1}^{P_{k}^{(i)}}2\kappa^{(i)}\Re\left\{\partial_{t}\tilde{Q}_{p, k}^{(i)}\right\} (\mathrm{d}_{\rho}E_{k})  &= 0,
\end{split}
\end{equation}
where we used the identities $2\Re\left\{\partial_{t}\tilde{Q}_{p, k}^{(i)}\right\} = \partial_{t}\tilde{Q}_{p, k}^{(i)} + \partial_{t}\tilde{Q}_{p, k}^{*(i)}$, and $\mathrm{d}_{\rho}E^{*}_{k} = \mathrm{d}_{\rho}E_{k}$.\par 
For a better readability, we will waive the symbol ``$\mathrm{d}^3r\,\mathrm{d}t$'' denoting the differential of the variable $(\mathbf{r}, t)$ in all following integral expressions. Integrating over space and time $\Omega \times I$, considering $\rho$ not to be time-dependent, and applying integration by parts in Eqs.~(\ref{Eq:MaxwellsDerivE}) and (\ref{Eq:MaxwellsDerivH}), while taking the imposed boundary conditions into account, leads to
\begin{equation} \label{Eq:Integr1}
\begin{split}
\int_{\Omega}\rlap{$\overbrace{\phantom{\left.\sum_{k=1}^{3}\varepsilon_{0}(\mathrm{d}_{\rho}\varepsilon_{\infty, k})\tilde{E}_k E_{k}\right|_0 ^T }}^{=\;0}$}\left.\sum_{k=1}^{3}\varepsilon_{0}(\mathrm{d}_{\rho}\varepsilon_{\infty, k})\tilde{E}_k E_{k}\right|_0 ^T 
&- \int_{\Omega}\int_{I}\sum_{k=1}^{3}\varepsilon_{0}(\mathrm{d}_{\rho}\varepsilon_{\infty, k})\partial_{t}\tilde{E}_k E_{k}\\
+\int_{\Omega}\rlap{$\overbrace{\phantom{\left.\sum_{k=1}^{3}\varepsilon_{0}\varepsilon_{\infty, k}\tilde{E}_k (\mathrm{d}_{\rho}E_{k})\right|_0 ^T}}^{=\;0}$}\left.\sum_{k=1}^{3}\varepsilon_{0}\varepsilon_{\infty, k}\tilde{E}_k (\mathrm{d}_{\rho}E_{k})\right|_0 ^T 
&- \int_{\Omega}\int_{I}\sum_{k=1}^{3}\varepsilon_{0}\varepsilon_{\infty, k}\partial_{t}\tilde{E}_k (\mathrm{d}_{\rho}E_{k})\\
&+ \int_{\Omega}\int_{I} \sum_{k=1}^{3}(\mathrm{d}_{\rho}\sigma_{k}) \tilde{E}_k E_{k}
+ \int_{\Omega}\int_{I} \sum_{k=1}^{3}\sigma_{k} \tilde{E}_k (\mathrm{d}_{\rho}E_{k})\\
+\int_{\Omega}\rlap{$\overbrace{\phantom{\left.\sum_{k=1}^{3}\sum_{i=1}^{2}\sum_{p=1}^{P_{k}^{(i)}}2(\mathrm{d}_{\rho}\kappa^{(i)})\tilde{E}_{k}\Re\left\{Q_{p, k}^{(i)}\right\}\right|_0 ^T}}^{=\;0}$}\left.\sum_{k=1}^{3}\sum_{i=1}^{2}\sum_{p=1}^{P_{k}^{(i)}}2(\mathrm{d}_{\rho}\kappa^{(i)})\tilde{E}_{k}\Re\left\{Q_{p, k}^{(i)}\right\}\right|_0 ^T 
&- \int_{\Omega}\int_{I} \sum_{k=1}^{3}\sum_{i=1}^{2}\sum_{p=1}^{P_{k}^{(i)}}2(\mathrm{d}_{\rho}\kappa^{(i)})\partial_{t}\tilde{E}_{k}\Re\left\{Q_{p, k}^{(i)}\right\}\\
+\int_{\Omega}\rlap{$\overbrace{\phantom{\left.\sum_{k=1}^{3}\sum_{i=1}^{2}\sum_{p=1}^{P_{k}^{(i)}}2\kappa^{(i)}\tilde{E}_{k}\Re\left\{(\mathrm{d}_{\rho}Q_{p, k}^{(i)})\right\}\right|_0 ^T}}^{=\;0}$}\left.\sum_{k=1}^{3}\sum_{i=1}^{2}\sum_{p=1}^{P_{k}^{(i)}}2\kappa^{(i)}\tilde{E}_{k}\Re\left\{(\mathrm{d}_{\rho}Q_{p, k}^{(i)})\right\}\right|_0 ^T 
&- \int_{\Omega}\int_{I}\sum_{k=1}^{3}  \sum_{i=1}^{2}\sum_{p=1}^{P_{k}^{(i)}}2\kappa^{(i)}\partial_{t}\tilde{E}_{k}\Re\left\{(\mathrm{d}_{\rho}Q_{p, k}^{(i)})\right\}\\
-\int_{\Omega}\rlap{$\overbrace{\phantom{\left.\sum_{k=1}^{3}\partial_{k}(\mathrm{d}_{\rho}\mathbf{H} \times \tilde{\mathbf{E}} )_{k}\right|_0 ^T}}^{=\;0}$}\left.\sum_{k=1}^{3}\partial_{k}(\mathrm{d}_{\rho}\mathbf{H} \times \tilde{\mathbf{E}} )_{k}\right|_0 ^T 
&- \int_{\Omega}\int_{I}\sum_{k=1}^{3} (\nabla \times \mathbf{\tilde{E}})_{k} (\mathrm{d}_{\rho} H_{k})\\
+\int_{\Omega}\rlap{$\overbrace{\phantom{\left.\sum_{k=1}^{3}\mu_0 \tilde{H_k} \mathrm{d}_{\rho}H_k\right|_0 ^T}}^{=\;0}$}\left.\sum_{k=1}^{3}\mu_0 \tilde{H_k} \mathrm{d}_{\rho}H_k\right|_0 ^T 
&- \int_{\Omega}\int_{I} \sum_{k=1}^{3} \mu_0 \partial_{t}\tilde{H_k} (\mathrm{d}_{\rho}H_k) \\
+\int_{\Omega}\rlap{$\overbrace{\phantom{\left.\sum_{k=1}^{3}\partial_{k}(\mathbf{\tilde{H}} \times \mathrm{d}_{\rho}\mathbf{E})_k\right|_0 ^T}}^{=\;0}$}\left.\sum_{k=1}^{3}\partial_{k}(\mathbf{\tilde{H}} \times \mathrm{d}_{\rho}\mathbf{E})_k\right|_0 ^T 
&+ \int_{\Omega}\int_{I} \sum_{k=1}^{3} (\nabla \times \mathbf{\tilde{H}})_{k}(\mathrm{d}_{\rho}E_k)\\[8pt]
&=0.
\end{split}
\end{equation}
We do the same  for Eq.~(\ref{Eq:Qsum}) and obtain:
\begin{equation} \label{Eq:Integr2}
\begin{split}
&\int_{\Omega}\rlap{$\overbrace{\phantom{\left.\sum_{k=1}^{3}\sum_{i=1}^{2}\sum_{p=1}^{P_{k}^{(i)}}\left(\frac{-\kappa^{(i)}}{\varepsilon_{0} c_{p, k}^{(i)}}\partial_{t}\tilde{Q}_{p, k}^{(i)}(\mathrm{d}_{\rho}Q_{p, k}^{(i)}) + \mathrm{c.c}\right)\right|_0 ^T}}^{=\;0}$}\left.\sum_{k=1}^{3}\sum_{i=1}^{2}\sum_{p=1}^{P_{k}^{(i)}}\left\{\frac{-\kappa^{(i)}}{\varepsilon_{0} c_{p, k}^{(i)}}\partial_{t}\tilde{Q}_{p, k}^{(i)}(\mathrm{d}_{\rho}Q_{p, k}^{(i)})+ \mathrm{c.c}\right\}\right|_0 ^T \\
+ &\int_{\Omega}\int_{I} \sum_{k=1}^{3} \sum_{i=1}^{2}\sum_{p=1}^{P_{k}^{(i)}}\left\{\frac{\kappa^{(i)}}{\varepsilon_{0} c_{p, k}^{(i)}}\partial^{2}_{t}\tilde{Q}_{p, k}^{(i)}(\mathrm{d}_{\rho}Q_{p, k}^{(i)}) + \frac{\kappa^{(i)} a_{p, k}^{(i)}}{\varepsilon_{0} c_{p, k}^{(i)}}\partial_{t}\tilde{Q}_{p, k}^{(i)} (\mathrm{d}_{\rho}Q_{p, k}^{(i)}) + \mathrm{c.c.}\right\}\\
+&\int_{\Omega}\int_{I} \sum_{k=1}^{3}\sum_{i=1}^{2}\sum_{p=1}^{P_{k}^{(i)}}2\kappa^{(i)}\Re\left\{\partial_{t}\tilde{Q}_{p, k}^{(i)}\right\} (\mathrm{d}_{\rho}E_{k})\\[8pt]
&=0.
\end{split}
\end{equation}
By adding Eqs.~(\ref{Eq:Integr1}) and~(\ref{Eq:Integr2}), and the integral of the functional derivative in Eq.~(\ref{Eq:ObjectiveDerivative}) over $\Omega \times I$, we obtain
\begin{equation}\label{Eq:integralRearanged}
\begin{split}
&\int_{\Omega}\int_{I}\sum_{k=1}^{3} \left\{\left( (\nabla \times \mathbf{\tilde{H}})_{k} - \varepsilon_{0}\varepsilon_{\infty, k}\partial_{t}\tilde{E}_k  + \sigma_{k} \tilde{E}_k + 2\sum_{i=1}^{2}\sum_{p=1}^{P_{k}^{(i)}}\kappa^{(i)}\Re\left\{\partial_{t}\tilde{Q}_{p, k}^{(i)}\right\} - \left(\frac{\delta F[\mathbf{E}]}{\delta \mathbf{E}}\right)_{k}\right)           
(\mathrm{d}_{\rho}E_k)\right\}\\
 +&\int_{\Omega}\int_{I} \sum_{k=1}^{3} \sum_{i=1}^{2}\sum_{p=1}^{P_{k}^{(i)}}\left\{\left(\frac{\kappa^{(i)}}{\varepsilon_{0} c_{p, k}^{(i)}}\partial^{2}_{t}\tilde{Q}_{p, k}^{(i)} + \frac{\kappa^{(i)} a_{p, k}^{(i)}}{\varepsilon_{0} c_{p, k}^{(i)}}\partial_{t}\tilde{Q}_{p, k}^{(i)} - \kappa^{(i)}\partial_{t}\tilde{E}_{k}\right)(\mathrm{d}_{\rho}Q_{p, k}^{(i)}) + \mathrm{c.c.}\right\}\\
 +&\int_{\Omega}\int_{I} \sum_{k=1}^{3}\left\{\left( - \mu_0 \partial_{t}\tilde{H_k} - (\nabla \times \mathbf{\tilde{E}})_{k} \right)(\mathrm{d}_{\rho}H_k)\right\}\\[8pt]
 -&\int_{\Omega}\int_{I}\frac{\delta F[\mathbf{E}]}{\delta \rho} + \int_{\Omega}\int_{I}\sum_{k=1}^{3}\left\{ \varepsilon_{0}(\mathrm{d}_{\rho}\varepsilon_{\infty, k})\partial_{t}\tilde{E}_k E_{k} - (\mathrm{d}_{\rho}\sigma_{k}) \tilde{E}_k E_{k} + 2\;\sum_{i=1}^{2}\sum_{p=1}^{P_{k}^{(i)}}(\mathrm{d}_{\rho}\kappa^{(i)})\partial_{t}\tilde{E}_{k}\Re\left\{Q_{p, k}^{(i)}\right\}\right\}\\[8pt]
 &=0.
\end{split}
\end{equation}
We note that the Eq.~(\ref{Eq:integralRearanged}) is satisfied, if the following equations hold $\forall (\mathbf{r},t) \in \Omega \times I$ and each spatial component $k \in \{1,2,3\}$:
\begin{subequations}\label{Eq:AlmostAdjoint}
  \begin{empheq}[]{align}
(\nabla \times \mathbf{\tilde{H}})_{k} - \varepsilon_{0}\varepsilon_{\infty, k}\partial_{t}\tilde{E}_k  + \sigma_{k} \tilde{E}_k + 2\sum_{i=1}^{2}\sum_{p=1}^{P_{k}^{(i)}}\kappa^{(i)}\Re\left\{\partial_{t}\tilde{Q}_{p, k}^{(i)}\right\}&= \left(\frac{\delta F[\mathbf{E}]}{\delta \mathbf{E}}\right)_{k},\\[1pt]
\text{For } i =1,2 \text{ and } \forall p \in 1, \ldots, P_k^{(i)}: \partial_{t}\tilde{Q}_{p, k}^{(i)}+a_{p, k}^{(i)}\tilde{Q}_{p, k}^{(i)}- \varepsilon_{0}c_{p, k}^{(i)}\tilde{E}_{k} &= 0,\\[8pt]
\mu_0 \partial_{t}\tilde{H_k} +(\nabla \times \mathbf{\tilde{E}})_{k} &=0,
\end{empheq}
\end{subequations}
and if $\forall \mathbf{r}\in \Omega$ the gradient of the objective $\nabla_{\rho}F[\mathbf{E}]$ defined  as
\begin{equation}\label{Eq:IntegralKernel}
\begin{split}
\nabla_{\rho}F[\mathbf{E}]:=\int_{I}\frac{\delta F[\mathbf{E}]}{\delta \rho},
\end{split}
\end{equation}
satisfies the equation
\begin{equation}\label{Eq:Gradients2}
\begin{split}
\nabla_{\rho}F[\mathbf{E}]=&\phantom{+}\int_{I}\sum_{k=1}^{3}\varepsilon_{0}(\mathrm{d}_{\rho}\varepsilon_{\infty, k})\partial_{t}\tilde{E}_k E_{k} \\
&-\int_{I}\sum_{k=1}^{3}(\mathrm{d}_{\rho}\sigma_{k}) \tilde{E}_k E_{k} \\
 &+\int_{I}\sum_{k=1}^{3}\sum_{i=1}^{2}\sum_{p=1}^{P_{k}^{(i)}}2(\mathrm{d}_{\rho}\kappa^{(i)})\partial_{t}\tilde{E}_{k}\Re\left\{Q_{p, k}^{(i)}\right\}.
\end{split}
\end{equation}
Now, we perform transformations of the fields in Eqs.~(\ref{Eq:AlmostAdjoint}) to obtain a system of Maxwell equations for the adjoint system. First, we reverse the time and change the sign of the magnetic field $\mathbf{\tilde{H}}$  and the currents $\tilde{Q}_{p, k}^{(i)}$ accordingly, i.e.
\begin{subequations}\label{Eq:Tranformations}
  \begin{empheq}[]{align}
  \mathbf{E}(t)&\rightarrow \mathbf{E}(\tau)\\
\mathbf{\tilde{H}}(t)&\rightarrow-\mathbf{\tilde{H}}(\tau),\\
\tilde{Q}_{p, k}^{(i)}(t)&\rightarrow-\tilde{Q}_{p, k}^{(i)}(\tau), \quad \text{for } k\in\{1,2,3\},\;i =1,2 \text{ and } \forall p \in 1, \ldots, P_k^{(i)}.
\end{empheq}
\end{subequations}
where $\tau:= T - t$ denotes the time-reversed variable. Furthermore, we require vanishing fields for $\tau=0$. If we now apply the chain rule for the time derivatives of all time reversed functions, we finally obtain the adjoint system which holds $\forall (\mathbf{r}, \tau) \in \Omega \times [0, T]$,
\begin{subequations}\label{Eq:AdjointSystem}
  \begin{empheq}[]{align}
-(\nabla \times \mathbf{\tilde{H}})_{k} + \varepsilon_{0}\varepsilon_{\infty, k}\partial_{\tau}\tilde{E}_k  + \sigma_{k} \tilde{E}_k + 2\sum_{i=1}^{2}\sum_{p=1}^{P_{k}^{(i)}}\kappa^{(i)}\Re\left\{\partial_{\tau}\tilde{Q}_{p, k}^{(i)}\right\}&= \left(\overleftarrow{\frac{\delta F[\mathbf{E}]}{\delta \mathbf{E}}}\right)_{k},\\[1pt]
\text{For } i =1,2 \text{ and } \forall p \in 1, \ldots, P_k^{(i)}: \partial_{\tau}\tilde{Q}_{p, k}^{(i)}-a_{p, k}^{(i)}\tilde{Q}_{p, k}^{(i)}- \varepsilon_{0}c_{p, k}^{(i)}\tilde{E}_{k} &= 0,\\[8pt]
\mu_0 \partial_{\tau}\tilde{H_k} +(\nabla \times \mathbf{\tilde{E}})_{k} &=0.
\end{empheq}
\end{subequations}
Here, the symbol ``$\overleftarrow{}$'' over the adjoint source term denotes the time-reversal transformation. 
Applying these transformation on the gradients in Eq.~(\ref{Eq:Gradients2}), leads to
\begin{equation}
\begin{split}
\nabla_{\rho}F[\mathbf{E}]=&-\int_{I}\sum_{k=1}^{3}\varepsilon_{0}(\mathrm{d}_{\rho}\varepsilon_{\infty, k})\partial_{\tau}\tilde{E}_k \overleftarrow{E}_{k} \\
&-\int_{I}\sum_{k=1}^{3}(\mathrm{d}_{\rho}\sigma_{k}) \tilde{E}_k \overleftarrow{E}_{k} \\
 &-\int_{I}\sum_{k=1}^{3}\sum_{i=1}^{2}\sum_{p=1}^{P_{k}^{(i)}}2(\mathrm{d}_{\rho}\kappa^{(i)})\partial_{\tau}\tilde{E}_{k}\Re\left\{\overleftarrow{Q}_{p, k}^{(i)}\right\}.
\end{split}
\end{equation}
This equation is equivalent to
\begin{equation}
\begin{split}
\nabla_{\rho}F[\mathbf{E}]=&-\int_{I}\sum_{k=1}^{3}\varepsilon_{0}(\mathrm{d}_{\rho}\varepsilon_{\infty, k})\partial_{\tau}\tilde{E}_k \overleftarrow{E}_{k} \\
&-\int_{I}\sum_{k=1}^{3}(\mathrm{d}_{\rho}\sigma_{k}) \tilde{E}_k \overleftarrow{E}_{k} \\
 &+\int_{I}\sum_{k=1}^{3}\sum_{i=1}^{2}\sum_{p=1}^{P_{k}^{(i)}}2(\mathrm{d}_{\rho}\kappa^{(i)})\tilde{E}_{k}\Re\left\{\partial_{\tau}\overleftarrow{Q}_{p, k}^{(i)}\right\},
\end{split}
\end{equation}
if we again apply integration by parts on the last term and taking the imposed boundary conditions into account.

\bibliography{supplement.bib}